\documentclass[a4,global,twocolumn]{svjour}

\usepackage{amsmath, amssymb}
\usepackage{pstricks}
\usepackage{graphicx}

\newcommand{\sgnC}[1]{\,\mathrm{sgn}_{\mathcal{C}}\!\left({#1}\right)}

\newcommand{\del}{\partial}
\newcommand{\n}{2}
\newcommand{\myi}{\mathrm{i}}
\renewcommand{\vector}[1]{\mathbf{#1}}

\DeclareMathOperator{\Trace}{Tr}
\newcommand{\Tr}[1]{\Trace\left[{#1}\right]}

\newcommand{\eq}[1]{(\ref{eq:#1})}
\newcommand{\Eq}[1]{Eq.~(\ref{eq:#1})}

\newcommand{\TwoEqs}[2]{Eqs.~(\ref{eq:#1}) and (\ref{eq:#2})}
\newcommand{\Fig}[1]{Fig.~\ref{fig:#1}}

\newcommand{\Sect}[1]{Sect.~\ref{sec:#1}}

\newcommand{\app}[1]{\ref{app:#1}}
\newcommand{\App}[1]{\mbox{App.\ \ref{app:#1}}}

\journalname{Applied physics B}

\begin{document}

\title{Far-from-equilibrium dynamics of an ultracold Fermi gas}
\author{Matthias~Kronenwett and Thomas~Gasenzer}
\institute{Institut f\"ur Theoretische Physik,
             Ruprecht-Karls-Universit\"at Heidelberg,
             Philosophenweg~16,
             69120~Heidelberg, Germany
           \and
           ExtreMe Matter Institute EMMI,
             GSI Helmholtzzentrum f\"ur Schwerionenforschung GmbH, 
             Planckstra\ss e~1, 
             64291~Darmstadt, Germany} 
\date{\today}

\maketitle

{ \sloppy 

\begin{abstract}
Nonequilibrium dynamics of an $\mathcal{N}$-fold spin-degenerate ultracold Fermi gas is described in terms of beyond-mean-field Kadanoff-Baym equations for correlation functions.
Using a nonperturbative expansion in powers of $1/\mathcal{N}$,
the equations are derived from the two-particle irreducible effective action in Schwinger-Keldysh formulation.
The definition of the nonperturbative approximation on the level of the effective action ensures vital conservation laws as, e.\,g., for the total energy and particle number.
As an example, the long-time evolution of a homogeneous, twofold spin-degenerate Fermi gas is studied in one spatial dimension after an initial preparation far from thermal equilibrium.
Analysis of the fluctuation-dissipation relation shows that, at low energies, the gas does not thermalise.
\end{abstract}


\section{Introduction}
\label{sec:Intro}

The preparation of ultracold atomic Bose and Fermi gases in various trapping environments allows to study important aspects of quantum many-body dynamics.
Many of the past experiments with ultracold gases can be approximately described by semi-classical approximations of the quantum many-body field equations such as by the Gross-Pitaevskii, Hartree-Fock-Bogoliubov, or Bardeen-Cooper-Schrieffer theories~\cite{Leggett2001a,Pitaevskii2003a}.
The description of the dynamics of many-body systems of sufficiently weakly interacting particles is usually based on perturbative approximations that rely on an expansion in powers of some dimensionless parameter that measures the binary interaction strength. 
These approximations are generically based on the smallness of statistical fluctuations.
In the limit of infinitely strong interactions, a number of approaches exists (in particular for systems in only one or two spatial dimensions) that allow dual descriptions in which approximations rely on the smallness of the inverse of the coupling strength~\cite{Korepin1997a,Glazman1996a}.

For intermediate interaction strengths, only a few approaches exist.
For such systems, quantum as well as strong classical fluctuations play in general an important r\^ole.
Prime examples are given by ultracold gases driven into the vicinity of Feshbach-Fano scattering resonances~\cite{Fano1935a,Fano1961a,Tiesinga1992a,Burnett1998a}, lattice-trapped gases in between the superfluid and Mott insulator regimes~\cite{Jaksch1998a,Bloch2004a,Morsch2006a}, and low-dimensional gases in regimes where the transverse confinement strongly affects the binary scattering dynamics of the particles~\cite{Pitaevskii2003a,Schmiedmayer2000a,Kinoshita2006a}.
Ultracold Fermi gases have been studied extensively in the vicinity of the BEC-BCS, i.\,e., superfluid-superconducting crossover~\cite{Regal2004a,Zwierlein2004a,Grimm2007a,Giorgini2008a,Inguscio2008a,Diehl:2009ma}, and currently attract increasing interest, e.\,g., in the context of Kondo phenomena in lattice environments~\cite{Gorshkov2009a}.

Taking into account higher-order classical and quantum fluctuations is important in describing the late-time behaviour of initially strongly perturbed as well as of continuously driven systems.
A quantitative description of the field dynamics in a closed system then generically requires nonperturbative approximations that need to be consistent with vital conservation laws such as the conservation of the total energy and the total particle number.
In this article, we present a self-consistent formulation of the nonequilibrium dynamics of an $\mathcal{N}$-component ultracold Fermi gas in terms of beyond-mean-field Kadanoff-Baym equations for correlation functions.
Dynamic equations for the two-point correlation functions are derived from the two-particle irreducible (2PI) effective action~\cite{Cornwall1974a,Luttinger1960a,Baym1962a} in Schwinger-Keldysh formulation~\cite{Schwinger1961a,Keldysh1964a}.
This approach extends the self-consistent mean-field formulation by including the effects of scattering between particles to the desired order of approximation and ensures the local conservation of the particle density-current vector as well as of the energy-momentum tensor.
Self-consistency is achieved by invoking approximations on the level of an effective action and obtaining time evolution equations by subsequent functional derivation with respect to the desired correlation functions.

Functional integral techniques based on the 2PI effective action enable an efficient embedding of summations of infinite series of perturbative processes.
In recent years, systematic nonperturbative expansions of the 2PI effective action to next-to-leading order in inverse powers of the number of field components~\cite{Hedin1965a} have allowed substantial progress~\cite{Berges:2001fi,Aarts:2002dj}.
After first successful applications of these nonperturbative expansions to the study of far-from-equilibrium dynamics as well as thermalization in relativistic bosonic~\cite{Berges:2001fi,Cooper:2002qd,Arrizabalaga:2004iw} and fermionic~\cite{Berges:2002wr,Berges:2004ce} theories, they have recently been employed in the context of ultracold bosonic quantum gases~\cite{Rey2004a,Gasenzer:2005ze,Temme2006a,Berges:2007ym,Branschadel:2008sk,Scheppach:2009wu}.
A related alternative approach based on renormalization-group-like flow equations was presented in Ref.~\cite{Gasenzer:2008zz,Gasenzer2008a}.
For introductory texts see, e.\,g., Refs.~\cite{Berges:2004yj,Gasenzer2009a}.

Considering the case of a degeneracy in the $\mathcal{N}$ spin degrees of freedom, we derive the 2PI effective action to next-to-leading order (NLO) in a nonperturbative expansion in powers of $1/\mathcal{N}$.
While the $1/\mathcal{N}$ approximation is entirely classical in leading order, classical fluctuations and corrections induced by quantum fluctuations, both of which are introduced by scattering, are included in NLO in a nonperturbative manner.
A similar $1/\mathcal{N}$ expansion can be derived for $SU(\mathcal{N})$ symmetric models of Kondo lattice systems~\cite{Gorshkov2009a}.

After introducing our approach in detail, we present in this paper, as an example, the long-time evolution of a homogeneous, twofold spin degenerate Fermi gas in one spatial dimension following a quench in the interaction strength.
Our results extend upon the work presented in \cite{Kronenwett2010a}.
The considered model is integrable and has as many conserved quantities as there are degrees of freedom \cite{Yang1967a}.
Hence, it is expected not to thermalise in general \cite{Rigol2007a,Manmana2007a,Gangardt2008a}.

Its low-energy properties can be approximated by a Tomonaga-Luttinger liquid (TLL) model \cite{Tomonaga1950a} which contains a linear free dispersion.
The resulting model is known to form a low-energy fixed point of the full interacting one-dimensional Fermi gas.
Owing to the quadratic form into which the TLL Hamiltonian can be transformed by introducing bosonic particle-hole operators, the occupation numbers of the resulting quasiparticle modes represent conserved quantities.
In \cite{Cazalilla2006a,Kennes2010a}, the long-time evolution of TLL fermion and coupled fermion-boson models after an interaction quench was found to approach a generalised Gibbs ensemble \cite{Rigol2007a}, accounting for the conserved quasiparticle numbers.
We point out that non-thermal stationary states have been found also in a number of other models, both integrable and non-integrable \cite{Calabrese2007a,Kollath2007a,Eckstein2008a,Moeckel2008a}.

Here, we consider the dynamic evolution described by the full interacting fermionic Hamiltonian, approaching the problem of equilibration from the high-energy end.
The Kadanoff-Baym dynamic equations in 2PI NLO $1/\mathcal{N}$ approximation we employ are generically considered to describe thermalisation \cite{Berges:2001fi,Cooper:2002qd,Arrizabalaga:2004iw,Berges:2002wr,Berges:2004ce,Rey2004a,Gasenzer:2005ze,Temme2006a,Berges:2007ym,Branschadel:2008sk}. 
While the conserved quantities of the considered 1D Fermi gas are not known explicitly, and while we do not expect to recover the TLL model at very low energies, we find, by analysing the fluctuation-dissipation theorem, that the correlation functions emerging at late times are incompatible with a thermal ensemble.

Our paper is organised as follows:
In Sect.~\ref{sec:2PIEA-Approach}, we define the fermionic model Lagrangian and summarise the most relevant aspects of the 2PI effective action approach to nonequilibrium dynamics.
Integro-differential Kadanoff-Baym or Schwinger-Dyson dynamic equations for the two-point Green functions are obtained from the effective action. 
In Sect.~\ref{sec:Nonpert}, we present different nonperturbative approximations of the 2PI effective action based both on a coupling expansion and on an expansion in powers of $1/\mathcal{N}$.
As an exemplary application demonstrating the power of the method, we study,  in Sect.~\ref{sec:Results1D}, the long-time evolution of a homogeneous one-dimensional Fermi gas with twofold spin degeneracy from an inital state that is far from thermal equilibrium.
Our conclusions are drawn in Sect.~\ref{sec:Concl}.
In the appendices, we review relevant properties of Gra\ss mann variables and two-time Green functions, and provide  a discussion of the conservation of total energy and particle number by the approximated 2PI effective action.

\section{Two-particle irreducible effective action approach to nonequilibrium dynamics}
\label{sec:2PIEA-Approach}

\subsection{The model Lagrangian}
\label{sec:oursystem}

We consider the dynamical evolution of an ultracold Fermi gas of atoms far from thermal equilibrium.
The atoms are assumed to be, internally, in $\mathcal{N}$ different hyperfine states.
Hence, in the language of quantum field theory, we study the dynamics of nonrelativistic complex fermionic fields $\hat\Psi_{\alpha}(\vector{x},t)$ obeying canonical anticommutation relations
\begin{equation}
  [\hat\Psi_{\alpha}(\vector{x},t),\hat\Psi^\dagger_{\beta}(\vector{y},t)]_{+}
  =\delta_{\alpha\beta}\delta(\vector{x}-\vector{y})\,,
\end{equation}
where $[\cdot,\cdot]_{+}$ denotes the anticommutator.
The indices $\alpha$ and $\beta$ count the $\mathcal{N}$ hyperfine ``spin'' states.
$s$-wave contact interactions between atoms in different hyperfine states are assumed, while Pauli's principle forbids $s$-wave collisions between fermions that are internally in the same state.
$p$-wave and higher-order partial-wave contributions are neglected. 
Our formalism, though, can be readily extended to more complicated interaction potentials.
In the contact potential approximation, the interactions in the channel characterised by the asymptotic hyperfine states $\alpha$ and $\beta$ are described by the potential $V_{\text{int},\alpha\beta}(\mathbf{x}-\mathbf{y},t) = g_{\alpha\beta}(t)\delta(\mathbf{x}-\mathbf{y})$ with a possibly time-dependent coupling strength $g_{\alpha\beta}(t)$.
In three spatial dimensions, the coupling strenght is related to the scattering length $a_{\alpha\beta}$ between states  $\alpha$ and $\beta$ by the relationship $g_{\alpha\beta}=4\pi a_{\alpha\beta}/m$.
Hence, the interaction Hamiltonian reads
\begin{equation}
  H_{\text{int}}(t)
  = \frac{g_{\alpha\beta}(t)}{2}\int_{\vector{x}}
    \hat\Psi^{\dagger}_{\alpha}(\vector{x})\hat\Psi^{\dagger}_{\beta}(\vector{x})
    \hat\Psi_{\beta}(\vector{x})\hat\Psi_{\alpha}(\vector{x})\,,
  \label{eq:N_species_V_int}
\end{equation}
where $\int_{\vector{x}}\equiv \int\text{d}^{d}{x}$ in $d$ spatial dimensions.

In this article, we will be concerned with a functional integral formulation of the dynamical field theory that involves functional integrations over complex Gra\ss mann-valued fields $\psi_{\alpha}(x)$ obeying
\begin{equation}
  [\psi_{\alpha}(x), \psi_{\beta}(y)]_{+}
  = [\psi_{\alpha}(x), \psi^{*}_{\beta}(y)]_{+}
  = 0
\end{equation}
for any combination of $\alpha$, $\beta$, $x$, and $y$,   
instead of a formulation in terms of equal-time anticommuting field operators.
Here, $x\equiv(t,\vector{x})\equiv(x_{0},\vector{x})$, etc., denote space-time coordinates and the asterisk complex conjugation.
Properties of Gra\ss mann variables relevant for our discussion can be found in App.~\ref{app:grassmannVariables}.

The Lagrangian for the $\mathcal{N}$-component ultracold Fermi gas with the above interactions reads 
\footnote{If not otherwise stated, we use natural units, with $\hbar=1$.}
\begin{align}
\label{eq:lagrangiandesityforcomplexfields}
\begin{split}
  &  L[\psi_{\alpha},\psi^*_{\alpha}] \\
  &= \int_{\vector{x}}
     \biggl\{
       \frac{\myi}{2}
       \Bigl[
         \psi^*_{\alpha}(x)\del_{x_{0}}\psi_{\alpha}(x)
         -[\del_{x_{0}}\psi^*_{\alpha}(x)]\psi_{\alpha}(x)
       \Bigr] \\
  &\hspace{3em}\mbox{}
       + \psi^*_{\alpha}(x)\frac{\vector{\nabla}^2}{2m}\psi_{\alpha}(x)
         - \psi^*_{\alpha}(x)V_{\text{ext},\alpha\beta}(x)\psi_{\beta}(x) \\
  &\hspace{3em}\mbox{}
       - \frac{{\lambda}_{\alpha\beta}}{2\mathcal{N}}
         \psi^*_{\alpha}(x)\psi^*_{\beta}(x)\psi_{\beta}(x)\psi_{\alpha}(x)
     \biggr\}\,,
\end{split}
\end{align}
where $\del_{x_{0}}$ denotes the partial derivative with respect to time, and $V_{\text{ext},\alpha\beta}(x)$ are possibly time-dependent trapping potentials or other external-field matrix elements coupling the hyperfine levels $\alpha$ and $\beta$.
Summations over repeated indices are implied.
A factor of $1/\mathcal{N}$ has been taken out of the couplings $\lambda_{\alpha\beta}=\mathcal{N} g_{\alpha\beta}$ in order to make the relative weight of the interaction to the quadratic terms in the Lagrangian invariant under a rescaling of $\mathcal{N}$.
This will be of use when considering the expansion in powers of $1/\mathcal{N}$ in \Sect{2PI1N}.

Three symmetries play an important r\^ole:
Local conservation of particle number implies that the Lagrangian density possesses a global $U(1)$ symmetry within each hyperfine subspace.
Moreoever, if the couplings between the hyperfine levels through both the external field $V_{\text{ext},\alpha\beta}$ and the interactions $\lambda_{\alpha\beta}$ are all equal to each other symmetry under rotations in the space of  hyperfine states.
Furthermore, for vanishing external fields and time-independent interactions, the Lagrangian is Galilei invariant, implying a locally conserved energy-momentum tensor.

For the ease of numerical implementation, we choose a field basis $\psi_{\alpha,i}$ where the index $i$ distinguishes real and imaginary parts of the quantum field,
\begin{align}
  \psi_{\alpha,1}(x)&\equiv\sqrt{\n}\,\text{Re}[\psi_{\alpha}(x)]\,,\\
  \psi_{\alpha,2}(x)&\equiv\sqrt{\n}\,\text{Im}[\psi_{\alpha}(x)]\,.
  \label{eq:realfieldrepresentation}
\end{align}
To simplify the notation, we include the hyperfine index $\alpha$ and the field index $i$ into a single index $a=(\alpha,i)$.
Sums over $a$ imply a sum over $\alpha\in\{1,\ldots,\mathcal{N}\}$ and one over $i\in\{1,2\}$. 

The Gra\ss mann action
\begin{equation}
  S[\psi]=\int_{x_{0}}L[\psi^{}_{\alpha},\psi^{*}_{\alpha}]\,,
\end{equation}
with $\int_{x_0}\equiv\int\mathrm{d}x_0$, associated with the Lagrangian \eq{lagrangiandesityforcomplexfields} reads
\begin{equation}
\label{eq:Sclass}
  S[\psi]
  =\frac{1}{2}\int_{xy}
   \overline\psi_{a}(x)\myi G^{-1}_{0,ab}(x,y) \psi_{b}(y)
   +S_{\text{int}}[\psi]\,.
\end{equation}
Here, the inverse free fermionic propagator is given by
\begin{equation}
  \myi G^{-1}_{0,ab}(x,y)
  = \delta(x-y)
    \bigl(
      \myi\tau_{ab}\del_{x_{0}}-\delta_{i_{a}i_{b}}
      H^{\mathrm{1B}}_{\alpha\beta}(x)
    \bigr)\,,
\label{eq:freeinversepropagator}
\end{equation}
with the one-body Hamiltonian
\begin{align}
  H^{\mathrm{1B}}_{\alpha\beta}(x)
  &=-\frac{\vector{\nabla}^{2}_{\vector{x}}}{2m}\delta_{\alpha\beta}+V_{\text{ext}, \alpha\beta}(x)\,.
\end{align}
Here, $\int_{x}\equiv\int\mathrm{d}x_0\,\int\mathrm{d}^d x$ denotes the integration over the region of space-time under consideration, $\delta_{ab}=\delta_{\alpha\beta}\delta_{i_{a}i_{b}}$, and
\begin{align}
  \tau_{ab}
  &\equiv-\delta_{\alpha\beta}\sigma^{2}_{i_{a}i_{b}}\,,
  &\sigma^{2}
  &\equiv\begin{pmatrix}0&-\myi\\ \myi&0\end{pmatrix}\,.
\label{eq:DefinitionOfTau}
\end{align}
Furthermore, $\delta(x-y)\equiv\delta(x_{0}-y_{0})\delta^{(d)}(\vector{x}-\vector{y})$ denotes the $(d+1)$-dimensional Dirac distribution and $\overline\psi_a(x)=\psi_b(x)\tau_{ba}$.
The interaction part $S_{\text{int}}[\psi]$  corresponding to the Lagrangian~\eqref{eq:lagrangiandesityforcomplexfields} is
\begin{equation}
  S_{\text{int}}[\psi]
  =-\frac{\lambda_{\alpha\beta}}{8\mathcal{N}}
   \int_{x}\overline\psi_{a}(x)\psi_{a}(x)\overline\psi_{b}(x)\psi_{b}(x)\,.
  \label{eq:interactionpartofS}
\end{equation}

Note that for a Fermi gas trapped in a lattice potential, the action, in the tight binding approximation, has the same form as in Eq.~\eq{Sclass}, and the free inverse propagator is given by
\begin{equation}
\begin{split}
  \myi G_{0,ab}^{-1}(x,y)
  &= \myi\tau_{ab} \partial_{x_0} \delta(x-y) \\
  &\quad\mbox{}
     - \delta_{ab} H^{\mathrm{1B}}_\mathrm{\alpha}(x,y) \delta(x_0-y_0)\,,
\end{split}
\end{equation}
where
\begin{equation}
  H^{\mathrm{1B}}_\mathrm{\alpha}(x,y)
  = -J \delta^{(d)}_{\langle \vector{n},\vector{m}\rangle}
    + \epsilon_{\alpha,n} \delta^{(d)}_{\vector{n}\vector{m}}
\end{equation}
is the one-body Hamiltonian. 
Here, $x=(x_0,\vector{n})$ and $y=(y_0,\vector{m})$ denote the lattice space-time coordinates, $\delta(x-y)=\delta^{(d)}_{\vector{n}\vector{m}}\delta(x_{0}-y_{0})$,
and $\delta^{(d)}_{\langle \vector{n},\vector{m}\rangle}=1$ if and only if, in the single-band approximation, $\vector{n}$ and $\vector{m}$ denote adjacent sites in the $d$-dimensional lattice; otherwise, $\delta^{(d)}_{\langle \vector{n},\vector{m}\rangle}=0$.
The site dependent energy $\epsilon_{\alpha,\vector{n}}$ describes, e.\,g., an additional external trapping potential for hyperfine mode $\alpha$, and we neglect spin mixing by this potential. 
All other previous and subsequent equations carry over to the lattice case when spatial integrals are replaced by sums over the lattice sites.
Extensions to the case of two or more internal electronic states and more specific symmetries in the hyperfine levels, as used, e.\,g., to describe Kondo lattice systems~\cite{Gorshkov2009a}, are straightforward and will be the subject of a subsequent publication.

\subsection{Nonequilibrium generating functional}
\label{sec:NEqGenFunc}

Knowing the time dependence of the correlation functions of a many-body system allows to derive the dynamics of physical observables.
These correlation functions can be obtained from the nonequilibrium generating functional $Z[K;\hat\rho_{D}]$,
\begin{equation}
  Z[K;\hat\rho_{D}]
    \equiv\Tr{\hat\rho_{D}(t_{0})
              \mathcal{T}_{\mathcal{C}}
              e^{\frac{\myi}{2}\int_{\mathcal{C},xy}\hat{\overline\Psi}_{a}(x)K_{ab}(x,y)\hat\Psi_{b}(y)}
             }\,,
\end{equation}
where $\hat\rho_{D}(t_{0})$ is the normalised density matrix of the system at the initial time $t_{0}$, and
  $\hat{\overline\Psi}_{a}(x) = \hat\Psi_{b}(x) \tau_{ba}$,
  $\hat{\Psi}_{a}(x)$
are field operators in the Heisenberg picture.
$\mathcal{C}=\mathcal{C}^+\cup\mathcal{C}^-$ indicates that the temporal integrals are taken along the closed (Schwinger-Keldysh) time path (CTP)~\cite{Schwinger1961a,Keldysh1964a} from the initial time $t_{0}$ to infinity (path $\mathcal{C}^+$) and back to $t_{0}$ ($\mathcal{C}^-$) such that
  $\int_{\mathcal{C},x} = \int_{\mathcal{C},x_{0}}\int_{\vector{x}}$
and
  $\int_{\mathcal{C},x_{0}} = \int_{\mathcal{C}^+}\mathrm{d}x_{0}
                              - \int_{\mathcal{C}^-}\mathrm{d}x_{0}$.
$\mathcal{T}_{\mathcal{C}}$ denotes time-ordering along the CTP, which implies that operators evaluated at later times stand to the left of those evaluated at earlier times.
The classical external two-point field
  $K_{ab}(x,y) = \tau_{bc} K_{cd}(y,x) \tau_{da}$
is introduced to allow for the generation of correlation functions of order $2n$,
\begin{equation}
\begin{split}
  &\langle
     \mathcal{T}_{\mathcal{C}}
     \hat{\overline\Psi}_{a_{1}}(x_{1}) \hat\Psi_{b_{1}}(y_{1})\cdots
     \hat{\overline\Psi}_{a_{n}}(x_{n}) \hat\Psi_{b_{n}}(y_{n})
    \rangle \\
  &=\mbox{}
     \frac{2^{n}}{Z}\left.\frac{\delta^{n} Z[K;\rho_{D}(t_{0})]}
    {\myi\delta K_{a_{1}b_{1}}(x_{1},y_{1})\cdots
     \myi\delta K_{a_{n}b_{n}}(x_{n},y_{n})}\right|_{K\equiv0}\,,
\end{split}
\label{eq:2npointGreen}
\end{equation}
where $\langle\cdot\rangle\equiv\Tr{\hat\rho_{D}\cdot}$.
Translated into a functional integral, the generating functional takes the form
\begin{equation}
\label{eq:genfunctional-w-inimatrix}
\begin{split}
  &Z[K;\rho_{D}(t_{0})]
   =\int\mathcal{D}\psi_{0}^+\mathcal{D}\psi_{0}^-
    \langle\phi_{0}^+|\hat\rho_{D}(t_{0})|\phi_{0}^-\rangle
  \\
  &\ \times\ 
   \int\mathcal{D}'\psi
   \,\exp\biggl\{\myi S_{\mathcal{C}}[\psi]
     +\frac{\myi}{2}
   \int_{\mathcal{C},xy} \overline\psi_{a}(x)K_{ab}(x,y)\psi_{b}(y)
     \biggr\}\,.
\end{split}
\end{equation}
Here, $|\phi_{0}^\pm\rangle$  are eigenstates of binary field operator products evaluated at the beginning and the end of the CTP, respectively,
  $\hat{\overline\Psi}_{a}(\vector{x},t_{0})
   \hat\Psi_{a}(\vector{x},t_{0})
   |\phi_{0}^\pm\rangle
   = -\myi\psi_{b}^{\pm}(\vector{x},t_{0})\tau_{ba}
     \psi_{a}^{\pm}(\vector{x},t_{0})
     |\phi_{0}^\pm\rangle$. 
The differential parts of the path integral measures are defined as
  $\mathcal{D}\psi_{0}^\pm
   = \prod_{\alpha,\vector{x}}
     \mathrm{d}\psi_{\alpha,1}^{\pm}(\vector{x},t_{0})
     \mathrm{d}\psi_{\alpha,2}^{\pm}(\vector{x},t_{0})$
and
  $\mathcal{D}'\psi
   = \prod_{\alpha,\mathbf{x},x_{0}\in\mathcal{C}\backslash\{t_{0}\}}
     \mathrm{d}\psi_{\alpha,1}(x)
     \mathrm{d}\psi_{\alpha,2}(x)$.
The action on the CTP is defined as 
\begin{equation}
  S_{\mathcal{C}}[\psi]
  =\int_{\mathcal{C},x_{0}}L[\psi_{\alpha},\psi^{*}_{\alpha}]\,.
\end{equation}

The most general initial density matrix can be parametrised as~\cite{Berges:2004yj,Gasenzer2009a}
\begin{equation}
  \langle \phi_{0}^{(+)} | \hat\rho_{D}(t_{0}) | \phi_{0}^{(-)} \rangle
  = \mathcal{N}^{}_{0} \exp\bigl[ \myi f_{\mathcal{C}}[\psi] \bigr]
\end{equation}
with normalization factor $\mathcal{N}^{}_{0}$ and a functional $f_{\mathcal{C}}[\psi]$ that can be expanded in powers of the fields,
\begin{align}
    f_{\mathcal{C}}[\psi]
    = \alpha^{(0)}_{}
      + \sum_{n=1}^{\infty} \frac{1}{n!}
        \int_{\mathcal{C},x^{1}\dotsb x^{n}}
       &\alpha^{(n)}_{a_1\dotsb a_n}(x^{1}_{},\dotsc,x^{n}_{}) \nonumber\\
    &\mbox{}\times
        \prod_{m=1}^{n} \psi_{a_m}^{}(x^{m}_{}) \,,
\end{align}
where the above boundary conditions are implied.
The cumulants $\alpha^{(n)}_{a_1\dotsb a_n}(x^{1}_{},\dotsc,x^{n}_{})$ are non-zero only at time $t^{}_{0}$ where the density matrix $\hat\rho_{D}(t_{0})$ is specified.
In the following, we will only consider Gaussian initial states, for which the cumulants vanish for all $n>2$ also at time $t_0$.
This allows to combine the initial density matrix with the external source field by defining the nonlocal source $R_{ab}^{}(x,y) = K_{ab}^{}(x,y) - \tau_{ac}\alpha^{(2)}_{cb}(x,y)$.
This allows to write the generating functional in the simpler form
\begin{equation}
\label{eq:stdformofnoneqgenfunctional}
\begin{split}
  Z[K]
  &= \int\mathcal{D}\psi\,
     \exp\biggl\{\myi S_{\mathcal{C}}[\psi]
     \\
  &\hspace{4em}\mbox{}
     + \frac{\myi}{2}
       \int_{\mathcal{C},xy} \overline\psi_{a}(x)R_{ab}(x,y)\psi_{b}(y)
     \biggr\}\,,
\end{split}
\end{equation}
where the measure also includes the fields at time $t_{0}$.

The connected two-point Green function%
\footnote{For fermionic fields, the two-point Green function $G$ is identical to the connected two-point function since the field expectation value always vanishes.
For higher $n$-point functions, there is, however, a distinction.} in the presence of the nonlocal source $R$ can be derived by functional differentiation,
\begin{equation}
  -\frac{1}{2} G_{ba}(y,x;R)
  = \frac{\delta W[R]}{\delta R_{ab}(x,y)} \,,
  \label{eq:G_from_W}
\end{equation}
of the Schwinger functional
\begin{equation}
\label{eq:generatingfunctional}
  W[R] = -\myi\ln Z[R]\,,
\end{equation}
which is the generating functional for $G$.
Later on, we use a vanishing external source $K=0$, and employ the notation, cf.\ \Eq{2npointGreen},
\begin{equation}
  G_{ab}(x,y)
  = \langle
      \mathcal{T}_{\mathcal{C}}
      \hat\Psi_{a}(x)
      \hat{\overline\Psi}_{b}(y)
    \rangle
  = G_{ab}(x,y;R[K=0]) \,.
  \label{eq:GreensFctForKZero}
\end{equation}
Symmetry properties of the two-point function $G$ as well as its decomposition into the physically relevant statistical and spectral correlation functions are discussed in detail in App.~\ref{app:2pGreenF}.

\subsection{2PI effective action}
\label{sec:2PIEA}

Directly evaluating the full quantum real-time path integral in \Eq{stdformofnoneqgenfunctional} is in general not feasible as the oscillating complex measure represents a variant of the sign problem.
One therefore needs to resort to analytical approaches in evaluating the dynamics in regimes where quantum fluctuations are relevant%
\footnote{This is in particular the case for long-time evolutions and where interactions become strong. If quantum fluctuations are small then the quantum part of the fluctuating fields can be integrated out leading to a classical path integral that can be computed using Monte Carlo techniques \cite{Berges:2007ym,Hillery:1983ms,Gardiner2004a,Blakie2008a,Polkovnikov2009a}.}.
 
The goal in deriving an effective action is similar to that of classical mechanics in the Lagrangian formulation:
The effective action $\varGamma[G]$ is defined such that it allows, by Hamilton's principle, to derive the dynamic equation for the correlation function $G$.
The action takes into account quantum effects, and the dynamic equations derived from it will obey crucial conservation laws corresponding to the symmetries of the underlying model Lagrangian.

The two-particle irreducible (2PI) effective action~\cite{Cornwall1974a,Luttinger1960a,Baym1962a} for the fermionic Lagrangian~\eq{lagrangiandesityforcomplexfields} is defined by a Legendre transform of $W[R]$ with respect to the source $R$~\footnote{Note that, as compared to the Bose case, there is no Legendre transform with respect to the one-point source $J_{a}(x)$ as there is no field expectation value, either.},
\begin{align}
  \varGamma[G]
  &= W[R]
     -\int_{xy}\frac{\delta W[R]}{\delta R_{ab}(x,y)}R_{ab}(x,y)\nonumber\\
  &= W[R] + \frac{1}{2}\Tr{GR}\,,
\end{align}
where we have used \Eq{G_from_W}, and where it is implied that \Eq{G_from_W} can be inverted to give $R$ as a function of $G$. 
The effective action $\varGamma[G]$ can be written as a series of terms represented by two-particle irreducible vacuum diagrams~\cite{Cornwall1974a},
\begin{equation}
  \varGamma[G]
  = -\frac{\myi}{2}
     \Tr{\ln(G^{-1})  
    + G_{0}^{-1}G}
    + \varGamma_{2}[G]+\text{const.}\,.
\label{eq:varGamma}
\end{equation}
The first term is a one-loop-type term derived from a saddle-point approximation of the path integral, where the trace denotes summation over all field and spin indices, and integration over spatial coordinates and over times along the CTP.
The constant term is irrelevant for the dynamics.
While the one-loop term remains within the mean-field approximation, scattering effects are contained in $\varGamma_{2}[G]$.
$\varGamma_{2}[G]$ can be written as the series of all vacuum, i.\,e., closed, 2PI diagrams constructed from the Green function $G$ and the bare vertices defined by the model Lagrangian~\cite{Cornwall1974a}.
A diagram is 2PI if it does not fall apart on opening two of its lines.
The specific expansion used in this work will be described in \Sect{Nonpert}.

\subsection{Dynamic equation for the Green function}
\label{sec:EOM}

Given the effective action $\varGamma$, the stationarity condition
\begin{equation}
  \frac{\delta\varGamma[G]}{\delta G_{ba}(y,x;R)}
  =\frac{1}{2}R_{ab}(x,y)
  \label{eq:thestationarycondition}
\end{equation}
determines $G$ for a given nonlocal source $R$.
For a given initial state, \Eq{thestationarycondition} represents the equation of motion for the Green function $G$.
Using \Eq{varGamma}, one finds the real-time Schwinger-Dyson- or Kadanoff-Baym-type equation
\begin{align}
  G_{ab}^{-1}(x,y;R)
  =G_{0,ab}^{-1}(x,y)-\myi R_{ab}(x,y)-\varSigma_{ab}(x,y;G)\,,
  \label{eq:eqnofmotionforvarDelta}
\end{align}
where
\begin{equation}
  \varSigma_{ab}(x,y;G)
  = -2\myi
    \frac{\delta\varGamma_{2}[G]}{\delta G_{ba}(y,x;R)}\,.
\end{equation}
denotes the one-particle irreducible (1PI) self-energy.
Convolving \Eq{eqnofmotionforvarDelta} with $G$ yields the dynamic equation for the two-point Green function,
\begin{align}
  &\int_{\mathcal{C},z}G_{0,ac}^{-1}(x,z)G_{cb}(z,y;R)
  = \delta_{\mathcal{C}}(x-y)\delta_{ab} \nonumber\\
  &\mbox{}\quad
    + \int_{\mathcal{C},z}
      \bigl(\varSigma_{ac}(x,z;G) + \myi R_{ac}(x,z)\bigr)
      G_{cb}(z,y;R) \,.
  \label{eq:finaleqnofmotionforvarDelta}
\end{align}
For a closed system (i.\,e., vanishing external source $K=0$), no explicit dependence on $R$ remains in the evolution equations since the term $\int_{\mathcal{C},z}\myi R_{ac}(x,z)G_{cb}(z,y;R)$ vanishes for $x_0\neq t_0$, i.\,e., the term becomes redundant for all times of interest.

Note that the free inverse propagator, \Eq{freeinversepropagator}, contains a first-order time derivative.
Since the free inverse propagator is otherwise diagonal in $(x-y)$, the integral on the left-hand side can be carried out, revealing the integro-differential structure of the dynamic equation~\eq{finaleqnofmotionforvarDelta}.
Despite \Eq{finaleqnofmotionforvarDelta} being exact, to be solved, it requires knowledge of the self-energy and therefore of the 2PI part $\varGamma_{2}$ of the effective action.
For practical computations, truncations of the series of 2PI diagrams are chosen as discussed in the following sections.

\section{Nonperturbative approximations of the 2PI effective action}
\label{sec:Nonpert}

In this section, we discuss different possible truncations of the expansion of $\varGamma_{2}$ in terms of two-particle irreducible (2PI) diagrams.
Figure~\ref{fig:2PI_Gamma2} shows the leading diagrams in the series ordered by the number of bare vertices per diagram.
The Green functions are represented by (blue) solid lines, the vertices by (black) dots.
\begin{figure}
  \begin{center}
    \includegraphics[width=.36\textwidth]{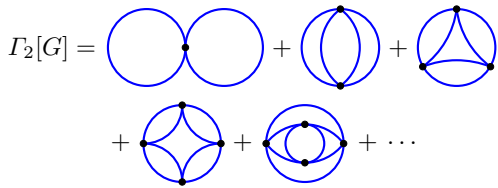}
  \end{center}
  \caption{ (Color online) 
     Diagrammatic expansion of the two-particle irreducible (2PI) part $\varGamma_{2}$ of the effective action~\eq{varGamma} in terms of 2PI graphs.
    (Blue) solid lines stand for the Green function $G$, and (black) dots for the interaction vertex $\lambda$.
    Explicitly shown are all diagrams that contain up to four vertices.
    All statistical factors and relative minus signs are omitted.
  }
  \label{fig:2PI_Gamma2}
\end{figure}

For the following discussion, it is convenient to separate out the local contributions to the proper self-energy,
\begin{equation}
\begin{split}
  \varSigma_{ab}(x,y;G)
   &= -\myi\varSigma^{(0)}_{ab}(x;G)\delta_{\mathcal{C}}(x-y) \\
   &\quad\mbox{}
      + \overline\varSigma_{ab}(x,y;G)\,,
\end{split}
\end{equation}
and include them together with the one-body Hamiltonian term of $G^{-1}_{0}$ in the matrix
\begin{equation}
\begin{split}\label{eq:Mab}
  M_{ab}(x,y;G)
  &=\delta_{\mathcal{C}}(x-y)
    \Bigl(
      \delta_{i_{a}i_{b}} H^{\mathrm{1B}}_{\alpha\beta}(x) \\
  &\hspace{6em}\mbox{}
      +\varSigma^{(0)}_{ab}(x;G)
    \Bigr)\,.
\end{split}
\end{equation}
The dynamic equation~\eqref{eq:finaleqnofmotionforvarDelta}, for $K=0$, now takes the compact form
\begin{multline}
\label{eq:finaleqnofmotionforFermi2ptFct}
  \myi\tau_{ac}\del_{x_{0}}
    G_{cb}(x,y)
  - \myi\delta_{ab}\delta_{\mathcal{C}}(x-y)\\
  = \int_{z}\left(M_{ac}(x,z;G)
    + \myi\overline\varSigma_{ac}(x,z;G)\right)G_{cb}(z,y)\,,
\end{multline}
where $G_{ab}(x,y)$ is defined in \Eq{GreensFctForKZero}.
Note that, neglecting $\overline\varSigma$, the dynamic equation reduces to a local differential equation in time, while the additional nonlocal contribution to the integrand introduces memory of the past evolution of $G$ into the equation.
Furthermore, it is convenient to reveal the specific structure of the point interaction vertex and to represent the vertex by a squiggly line.
The interaction part of the considered Lagrangian density~\eq{lagrangiandesityforcomplexfields} requires that spins are conserved at each end of the vertex.
Furthermore, it does not allow the Green functions in two adjacent loops to correspond to the same hyperfine quantum number since $\lambda_{\alpha\alpha}=0$.
The three possible index contractions at each vertex are depicted in \Fig{vertex}.
\begin{figure}
  \begin{center}
    \includegraphics[width=.37\textwidth]{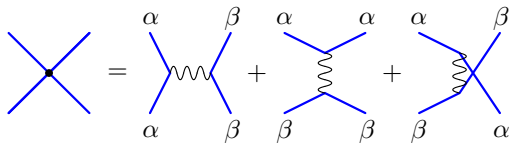}
  \end{center}
  \caption{ (Color online) 
    Decomposition of the bare vertex $\lambda$ (black dot) as a sum of the three possible spin-index contractions.
    The point interaction vertex $\lambda_{\alpha\beta}$ is represented by a squiggly line at each end of which spin and field indices are conserved and summed over.
  }
  \label{fig:vertex}
\end{figure}

\subsection{Coupling expansion}
\label{sec:Loop}

The expansion of $\varGamma_{2}$ in terms of 2PI diagrams can be ordered as a power series in the bare coupling constant $g$.
Note, that this, in the strict sense, does not constitute a perturbative expansion in powers of $g$ since the Green function $G$ entering the diagrams is determined self-consistently by \Eq{finaleqnofmotionforvarDelta}.
Thus, each diagram contains contributions up to arbitrarily high powers in the coupling.
Nevertheless, if the solution of the dynamic equations exhibits $G$ to be a bounded function then the coupling expansion of $\varGamma_{2}$ can be effectively viewed to be perturbative.

\subsubsection{Hartree-Fock-Bogoliubov approximation}
\label{sec:HFB}

As a first step, we recover the dynamic equations in the Hartree-Fock-Bogoliubov (HFB) approximation.
Retaining only the lowest-order diagram of the coupling expansion of $\varGamma_{2}$, that is, the double-bubble contribution shown in \Fig{HFBBAnd2ndOrder}, is known as the Hartree-Fock-Bogoliubov approximation.
\begin{figure}
  \begin{center}
    \includegraphics[width=.39\textwidth]{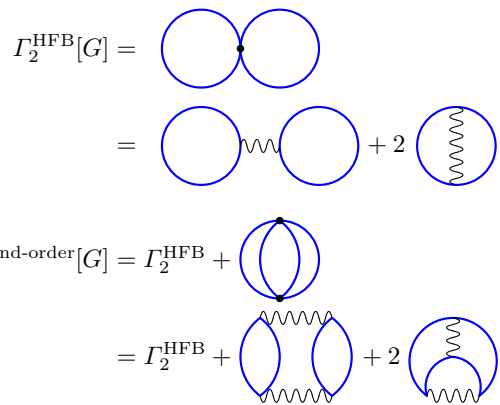}
  \end{center}
  \caption{ (Color online) 
    The first-order (Hartree-Fock-Bogoliubov) and second-order coupling expansions of $\varGamma_{2}$.
    The factors of 2 arise due to the different possible spin index contractions depicted in \Fig{vertex}.
    All other statistical factors and relative minus signs appearing because of the Gra\ss mann algebra properties are omitted.
  }
  \label{fig:HFBBAnd2ndOrder}
\end{figure}
Due to the structure of the vertex, \Eq{interactionpartofS}, this contribution consists of two qualitatively different diagrams and reads
\begin{equation}
\begin{split}\label{eq:HFBBcontribution2varGamma2ab}
  \varGamma^{\text{HFB}}_{2}[G]
  &= -\frac{\lambda_{\alpha\beta}}{8\mathcal{N}}
      \int_{x}
      \biggl(
        G_{aa}(x,x)G_{bb}(x,x) \\
  &\hspace{6em}\mbox{}
        - 2G_{ab}(x,x)G_{ba}(x,x)
      \biggr)\,,
\end{split}
\end{equation}
where it is summed over $a=(\alpha,i_a)$ and $b=(\beta,i_b)$.
The self-energy derived from \Eq{HFBBcontribution2varGamma2ab} is
\begin{align}
  \varSigma^{\text{HFB}}_{ab}(x,y;G)
  &= -\myi\varSigma^{\text{HFB}(0)}_{ab}(x;G)\delta_{\mathcal{C}}(x-y)\,,
\\
\begin{split} \label{eq:HFBB_selfenergy}
  \varSigma^{\text{HFB}(0)}_{ab}(x;G)
  &= -\frac{\lambda_{\alpha\gamma}}{2\mathcal{N}}
     \Big(
       \delta_{ab} G_{cc}(x,x)\\
  &\hspace{4.7em}\mbox{}
       - 2\delta_{\gamma\beta} G_{ab}(x,x)
     \Big)\,,
\end{split}
\end{align}
where it is summed over $c=(\gamma,i_c)$.
Hence, the HFB self-energy reduces to its local part $\varSigma^{\text{HFB}(0)}$.

We employ the decomposition of the full Green function $G$ into the statistical correlation function $F$ and the spectral function $\rho$,
\begin{equation}
  G_{ab}(x,y)
  = F_{ab}(x,y) - \frac{\myi}{2} \rho_{ab}(x,y) \sgnC{x_{0}-y_{0}}\,,
  \label{eq:decomposiontID4G}
\end{equation}
which are defined in Eqs.~\eq{definitionOfF} and \eq{definitionOfRho}, respectively, and discussed in more detail in App.~\app{2pGreenF}.
Inserting this decomposition into the dynamic equation for $G$ and using \Eq{dx0rho}, we find the time-dependent Hartree-Fock-Bogoliubov equations for $\rho$ and $F$,
\begin{align}
  \left[
  \myi\tau_{ac}\del_{x_{0}}
    -M^{\text{HFB}}_{ac}(x;G)
  \right]
  F_{cb}(x,y)
  &= 0\,,
\label{eq:dynamicHFBEqn4F}\\
  \left[
    \myi\tau_{ac}\del_{x_{0}}
    -M^{\text{HFB}}_{ac}(x;G)
  \right]
  \rho_{cb}(x,y)
  &= 0\,,
\label{eq:dynamicHFBEqn4rho}
\end{align}
where $M^{\text{HFB}}(x;G) $ is obtained by \Eq{Mab}, with $\varSigma^{(0)}(x;G) = \varSigma^{\text{HFB}(0)}(x;G)$ as given in \Eq{HFBB_selfenergy}.
As the nonlocal self-energy contribution vanishes in the HFB approximation, the dynamic equations for $F$ and $\rho$ decouple.
At equal times, $x_{0}=y_{0}$, the spectral function $\rho$ is fixed by the anticommutation relations \eq{equaltimeanticommutator}. 
Hence, one finds that the single-particle density matrix
  $n_{\alpha\beta}(\vector{x},\vector{y},t)
   = \langle
       \hat\Psi_{\alpha}^{\dagger}(\vector{x})
       \hat\Psi_{\beta}(\vector{y})
     \rangle_t$
is solely determined by the statistical correlation function:
\begin{equation}
\begin{split} \label{eq:SPartDensityMatrixintermsofF}
  &  n_{\alpha\beta}(\vector{x},\vector{y},t)
     - \frac{1}{2} \delta_{\alpha\beta} \delta(\vector{x}-\vector{y}) \\
  &= -\frac{1}{2}
      \delta_{i_{a}1}\delta_{i_{b}1}
     \Bigl(
       F_{ab}(\vector{x},\vector{y},t)
       + F_{\bar a\bar b}(\vector{x},\vector{y},t) \\
  &\hspace{6.5em}\mbox{}
       +\myi
        \bigl(
          F_{a\bar b}(\vector{x},\vector{y},t)
          - F_{\bar ab}(\vector{x},\vector{y},t)
        \bigr)
     \Bigr)\,,
\end{split}
\end{equation}
where ${\bar a}=(\alpha,3-i_a)$, and it is summed over $i_{a}$ and $i_{b}$.
This includes the density of particles in mode $\alpha$ at point $(\vector{x},t)$, $n_{\alpha}(\vector{x},t)\equiv n_{\alpha\alpha}(\vector{x},\vector{x},t)$. 
Moreover, the anomalous density matrix or pair function
  $m_{\alpha\beta}(\vector{x},\vector{y},t) 
   = \langle
       \hat\Psi_{\alpha}(\vector{x})
       \hat\Psi_{\beta}(\vector{y})
     \rangle_t$
is given as
\begin{equation}
\begin{split} \label{eq:PairFunctionintermsofF}
  &m_{\alpha\beta}(\vector{x},\vector{y},t) \\
  &\mbox{}
   = -\frac{1}{2}
      \delta_{i_{a}1}\delta_{i_{b}1}
     \Bigl(
       F_{ab}(\vector{x},\vector{y},t)
       - F_{\bar a\bar b}(\vector{x},\vector{y},t) \\
  &\hspace{6.5em}\mbox{}
       +\myi
        \bigl(
           F_{a\bar b}(\vector{x},\vector{y},t)
           + F_{\bar ab}(\vector{x},\vector{y},t)
        \bigr)
      \Bigr)\,,
\end{split}
\end{equation}
where it is summed over $i_{a}$ and $i_{b}$.
Adding \Eq{dynamicHFBEqn4F} and its transpose, one obtains the set of coupled HFB equations for $n_{\alpha\beta}(\vector{x},\vector{y},t)$ and $m_{\alpha\beta}(\vector{x},\vector{y},t)$:
\begin{align}
\begin{split} \label{eq:HFBEqForN}
  &  \bigl( \myi\partial_{t} + H_{\alpha\alpha}(x) - H_{\beta\beta}(y) \bigr)
     {\tilde n}_{\alpha\beta}(\vector{x}, \vector{y}, t)
  \\
  &= \biggl\{
       \frac{2 \lambda_{\alpha\gamma}}{\mathcal{N}}
       \Bigl(
         - {\tilde n}_{\gamma\gamma}(\vector{x}, \vector{x}, t)
           {\tilde n}_{\alpha\beta}(\vector{x}, \vector{y}, t)
  \\
  &\hspace{4.8em}\mbox{}
         + m_{\alpha\gamma}^*(\vector{x}, \vector{x}, t)
           m_{\gamma\beta}(\vector{x}, \vector{y}, t)
  \\
  &\hspace{4.8em}\mbox{}
         + {\tilde n}_{\alpha\gamma}(\vector{x}, \vector{x}, t)
           {\tilde n}_{\gamma\beta}(\vector{x}, \vector{y}, t)
       \Bigr)
     \biggr\}
  \\
  &\quad\mbox{}
     - \bigl\{
         (\alpha,\vector{x})\leftrightarrow(\beta,\vector{y})
       \bigr\}^*\,,
\end{split} \\
\begin{split}  \label{eq:HFBEqForM}
  &  \bigl( \myi\partial_{t} - H_{\alpha\alpha}(x) - H_{\beta\beta}(y) \bigr)
     m_{\alpha\beta}(\vector{x}, \vector{y}, t)
  \\
  &= \biggl\{
       \frac{2\lambda_{\alpha\gamma}}{\mathcal{N}}
       \Bigl(
         {\tilde n}_{\gamma\gamma}(\vector{x}, \vector{x}, t)
         m_{\alpha\beta}(\vector{x}, \vector{y}, t)
  \\
  &\hspace{4.8em}\mbox{}
         - {\tilde n}_{\alpha\gamma}^*(\vector{x}, \vector{x}, t)
           m_{\gamma\beta}(\vector{x}, \vector{y}, t)
  \\
  &\hspace{4.8em}\mbox{}
         - m_{\alpha\gamma}(\vector{x}, \vector{x}, t)
           {\tilde n}_{\gamma\beta}(\vector{x}, \vector{y}, t)
       \Bigr)
     \biggr\}
  \\
  &\quad\mbox{}
     - \bigl\{
         (\alpha,\vector{x})\leftrightarrow(\beta,\vector{y})
       \bigr\}\,,
\end{split}
\end{align}
where it is summed over $\gamma$, and
  ${\tilde n}_{\alpha\beta}(\vector{x},\vector{y},t)
   \equiv n_{\alpha\beta}(\vector{x},\vector{y},t)
          - \delta_{\alpha\beta}\delta(\vector{x}-\vector{y})/2$.
The last term in \Eq{HFBEqForM} (\Eq{HFBEqForN}) denotes (the complex conjugate of) the first term
in curly brackets with $\alpha$ and $\beta$, and $\vector{x}$ and $\vector{y}$ interchanged.
Equivalently, the HFB equations can be derived using the Ehrenfest theorem, i.\,e.,
  $\myi\partial_{t} {n}_{\alpha\beta}(\vector{x}, \vector{y}, t)
   = -\langle
        [H, \hat\Psi_{\alpha}^{\dagger}(\vector{x})
            \hat\Psi_{\beta}(\vector{y})]_{-}
      \rangle$
and similarly for $m_{\alpha\beta}(\vector{x}, \vector{y}, t)$, and a Weyl ordered Hamiltonian $H$.
When evaluating the expectation values, the Hartree-Fock-Bogoliubov approximation then consists of neglecting all joint cumulants higher than second order%
\footnote{For related discussions in the context of cold gases, see, e.\,g., Refs.~\cite{Gasenzer:2005ze,Proukakis1998a,Holland2001b,Kohler2002a,Naidon2003a}.}.

In this paper we will consider, as a concrete example, a homogeneous one-dimensional Fermi gas ``on a ring'', i.\,e., in a finite box with periodic boundary conditions.
In this case, the correlation functions only depend on the spatial relative coordinate $\vector{x}-\vector{y}$.
In this case, Eqs.~\eq{dynamicHFBEqn4F} and \eq{dynamicHFBEqn4rho} are conveniently solved in momentum space.
The solutions are
\begin{align}
\begin{split}  \label{eq:SolHFBEqn4F}
  F_{ab}(\vector{p},t)
  &= \bigl(
       \exp[-\myi\tau M^{\text{HFB}}(\vector{p}) t]
     \bigr)_{ac}\,
     F_{cb}(\vector{p},t)\,,
\end{split}\\
\begin{split}  \label{eq:SolHFBEqn4rho}
  \rho_{ab}(\vector{p},t)
  &= \bigl(
       \exp[-\myi\tau M^{\text{HFB}}(\vector{p}) t]
     \bigr)_{ac}\,
     \rho_{cb}(\vector{p},t)\,,
\end{split}
\end{align}
with
\begin{equation}
\begin{split}
  M_{ab}^{\text{HFB}}(\vector{p})
  &= \delta_{ab}
     \left(
       \frac{\vector{p}^2}{\n m} 
       - \frac{\lambda_{\alpha\gamma}}{2\mathcal{N}}
         \int_{\vector{q}}
         F_{cc}(\vector{q},t)
     \right)\\
  &\quad\mbox{}
       + \frac{\lambda_{\alpha\beta}}{\mathcal{N}}
         \int_{\vector{q}}
         F_{ab}(\vector{q},t)\,,
\end{split}
\end{equation}
where $\int_{\vector{q}}=(2\pi)^{-d}\int\mathrm{d}^{(d)}q$, and it is summed over $c=(\gamma, i_c)$.
Inserting the solution for $F$ into
  $n_{\alpha}(\vector{p},t)
   = [1
     - F_{(\alpha,1)(\alpha,1)}(\vector{p},t)
        - F_{(\alpha,2)(\alpha,2)}(\vector{p},t)]/\n$,
we recover that the HFB equations, which exclude scattering, leave all momentum-mode occupation numbers invariant.

\subsubsection{Second-order coupling approximation}
\label{sec:2ndloop}

Beyond the mean-field Hartree-Fock-Bogoliubov contribution to $\varGamma_{2}$, we now additionally take into account the second-order diagram shown in \Fig{HFBBAnd2ndOrder} containing two bare couplings:
\begin{equation}
\begin{split}
  &\varGamma_{2}^\text{2nd}[G]\\
  &= \varGamma_{2}^\text{HFB}[G]
     + \frac{\myi \lambda_{\alpha\beta} \lambda_{\gamma\delta}}{16\mathcal{N}^2}
       \int_{xy}
       G_{bc}(x,y) G_{cb}(y,x)
  \\
  &\hspace{11em}\mbox{}\times
       G_{da}(x,y) G_{ad}(y,x)
  \\
  &\hspace{4.5em}\mbox{}
  - \frac{\myi \lambda_{\alpha\beta} \lambda_{\gamma\delta}}{8\mathcal{N}^2}
       \int_{xy}
       G_{ad}(x,y) G_{db}(y,x)
  \\
  &\hspace{11em}\mbox{}\times
       G_{bc}(x,y) G_{ca}(y,x)\,,
  \label{eq:Gamma2-2nd-loop}
\end{split}
\end{equation}
where it is summed over $a=(\alpha, i_a)$, $b=(\beta, i_b)$, $c=(\gamma, i_c)$, and $d=(\delta, i_d)$.
Taking the functional derivative with respect to $G$, the additional term yields the nonlocal self-energy
\begin{equation}
\begin{split}
  \overline\varSigma_{ab}(x,y)
  &= \frac{\lambda_{\alpha\gamma} \lambda_{\delta\beta}}{2\mathcal{N}^2}
       \varPi_{\gamma\delta}(x,y) G_{ab}(x,y) \\
  &\quad\mbox{}
     - \frac{\lambda_{\alpha\gamma} \lambda_{\delta\beta}}{\mathcal{N}^2}
       G_{ad}(x,y) G_{dc}(y,x) G_{cb}(x,y)
  \label{eq:SigmaNonLocal-2nd-loop}
\end{split}
\end{equation}
with
\begin{equation}
  \varPi_{\alpha\beta}(x,y)
  = G_{(\alpha,i)(\beta,j)}(x,y)
    G_{(\beta,j)(\alpha,i)}(y,x)\,,
\label{eq:definitionOfPi}
\end{equation}
where sums over $\gamma$, $\delta$, $i_{c}$, and $i_{d}$ are implied in \Eq{SigmaNonLocal-2nd-loop}, and over $i$ and $j$ in \Eq{definitionOfPi}.

In order to derive the contributions to the dynamic equations for $F$ and $\rho$, one decomposes $\overline\varSigma$ and $\varPi$ into their statistical and spectral parts,
\begin{align}
\label{eq:decompositionSigmaFrho}
  \overline\varSigma_{ab}(x,y)
  &= \overline\varSigma^F_{ab}(x,y)
     - \frac{\myi}{2} \overline\varSigma^\rho_{ab}(x,y) \sgnC{x_{0}-y_{0}}\,,
\\
  \varPi_{\alpha\beta}(x,y)
  &= \varPi^F_{\alpha\beta}(x,y)
     - \frac{\myi}{2} \varPi^\rho_{\alpha\beta}(x,y) \sgnC{x_{0}-y_{0}}\,,
\end{align}
which gives
\begin{align}
\begin{split}\label{eq:nonlocalSigma2ndLoopF}
  \overline\varSigma^{F}_{ab}(x,y)
  &= \frac{\lambda_{\alpha\gamma} \lambda_{\delta\beta}}{2\mathcal{N}^2}
     \biggl(
       \varPi^{F}_{\gamma\delta}(x,y)
       F_{ab}(x,y)\\
  &\hspace{5.5em}\mbox{} 
       - \frac{1}{4}
         \varPi^{\rho}_{\gamma\delta}(x,y)
         \rho_{ab}(x,y)\\
  &\hspace{5.5em}\mbox{} 
       - 2P^{F}_{a\delta c}(x,y)
       F_{cb}(x,y)\\
  &\hspace{5.5em}\mbox{} 
       + \frac{1}{2}
         P^{\rho}_{a\delta c}(x,y)
         \rho_{cb}(x,y)
     \biggr)\,, 
\end{split}\\
\begin{split}\label{eq:nonlocalSigma2ndLoopRho}
  \overline\varSigma^{\rho}_{ab}(x,y)
  &= \frac{\lambda_{\alpha\gamma} \lambda_{\delta\beta}}{2\mathcal{N}^2}
     \biggl(
       \varPi^{F}_{\gamma\delta}(x,y)
       \rho_{ab}(x,y)\\
  &\hspace{5.5em}\mbox{} 
       + \varPi^{\rho}_{\gamma\delta}(x,y)
         F_{ab}(x,y)\\
  &\hspace{5.5em}\mbox{} 
       - 2P^{F}_{a\delta c}(x,y)
         \rho_{cb}(x,y)\\
  &\hspace{5.5em}\mbox{} 
       - 2P^{\rho}_{a\delta c}(x,y)
         F_{cb}(x,y)
     \biggr)\,,
\end{split}
\end{align}  
with
\begin{align}
\begin{split}\label{eq:definitionOfPF}
  P^{F}_{a\delta c}(x,y)
  &= F_{a(\delta, i)}(x,y)
     F_{(\delta, i)c}(y,x)\\
  &\quad\mbox{}
     + \frac{1}{4}
       \rho_{a(\delta, i)}(x,y)
       \rho_{(\delta, i)c}(y,x) \,,
\end{split}\\
\begin{split}
  P^{\rho}_{a\delta c}(x,y)
  &= \rho_{a(\delta, i)}(x,y)
     F_{(\delta, i)c}(y,x)\\
  &\quad\mbox{}
     - F_{a(\delta, i)}(x,y)
       \rho_{(\delta, i)c}(y,x) \,,
\end{split}\\
\begin{split}\label{eq:definitionOfPiF}
  \varPi^{F}_{\alpha\beta}(x,y)
  &= F_{(\alpha,i)(\beta,j)}(x,y)
     F_{(\beta,j)(\alpha,i)}(y,x)\\
  &\quad\mbox{}
     + \frac{1}{4}
       \rho_{(\alpha,i)(\beta,j)}(x,y)
       \rho_{(\beta,j)(\alpha,i)}(y,x) \,,
\end{split}\\
\begin{split}\label{eq:definitionOfPiRho}
  \varPi^{\rho}_{\alpha\beta}(x,y)
  &= \rho_{(\alpha,i)(\beta,j)}(x,y)
     F_{(\beta,j)(\alpha,i)}(y,x)\\
  &\quad\mbox{}
     - F_{(\alpha,i)(\beta,j)}(x,y)
       \rho_{(\beta,j)(\alpha,i)}(y,x) \,,
\end{split}
\end{align}
$a=(\alpha,i_a)$, $b=(\beta,i_b)$, and $c=(\gamma,i_c)$.
On the right-hand sides of Eqs.~\eqref{eq:nonlocalSigma2ndLoopF}--\eqref{eq:definitionOfPiRho}, it is summed over indices not appearing on the respective left-hand sides. 
Finally, inserting Eqs.~\eqref{eq:decomposiontID4G} and \eqref{eq:decompositionSigmaFrho} into \Eq{finaleqnofmotionforFermi2ptFct}, one finds the dynamic equations for $F$ and $\rho$:
\begin{align}
  &\bigl[\myi
    \tau_{ac}\del_{x_{0}}
    -M^{\text{HFB}}_{ac}(x;G)
  \bigr]
  F_{cb}(x,y)\nonumber\\
  &=\int^{x_{0}}_{t_0}\mathrm{d}z\,\overline\varSigma^{\rho}_{ac}(x,z;G)F_{cb}(z,y)\nonumber\\
  &\quad\mbox{} -\int^{y_{0}}_{t_0}\mathrm{d}z\,\overline\varSigma^{F}_{ac}(x,z;G)\rho_{cb}(z,y)\,,
  \label{eq:dynamicEqn4F}\\
  &\bigl[\myi
    \tau_{ac}\del_{x_{0}}
    -M^{\text{HFB}}_{ac}(x;G)
  \bigr]
  \rho_{cb}(x,y)\nonumber\\
  &=\int^{x_{0}}_{y_{0}}\mathrm{d}z\,\overline\varSigma^{\rho}_{ac}(x,z;G)\rho_{cb}(z,y)\,,
  \label{eq:dynamicEqn4rho}
\end{align}
where $\int^{t'}_{t}\mathrm{d}x \equiv \int^{t'}_{t}\mathrm{d}x_{0}\int_{\vector{x}}$.
In deriving these equations, the decomposition of $G$ and $\overline\varSigma$ into statistical and spectral parts allows to rewrite time integrals over the CTP into simple time integrals.
Hence, the right-hand sides introduce scattering effects in form of memory integrals that render the equations non-Markovian.
The form of Eqs.~\eq{dynamicEqn4F} and \eq{dynamicEqn4rho} is independent of the order of approximation chosen for $\varGamma_{2}$.
In the second-order coupling approximation introduced above, they form a closed set of integro-differential dynamic equations.

From the dynamic equations~\eqref{eq:dynamicEqn4F} and~\eqref{eq:dynamicEqn4rho}, standard quantum kinetic (Boltzmann) equations can be derived for the mode occupation numbers $n_{\alpha}(\vector p,t)$.
This is achieved by a transformation to Wigner space, neglecting initial-time and non-Markovian effects in a gradient expansion with respect to the absolute time $T=(x_{0}+y_{0})/2$, and making a quasiparticle ansatz -- cf., e.\,g., Refs.~\cite{Berges:2004yj,Rey2005a} for details.
The relevance of non-Markovian and initial-time effects provided by the full dynamic equations has been discussed, for bosonic theories, in Refs.~\cite{Berges:2005md,Branschadel:2008sk}.
For concise discussions of kinetic equations, we refer to Refs.~\cite{KadanoffBaym1962,Bonitz1998}.

Following the procedure outlined above, higher-order coupling approximations can be derived straightforwardly.
The resulting equations of motion for the two-point correlation functions have the form~\eqref{eq:dynamicEqn4F} and~\eqref{eq:dynamicEqn4rho} with modified proper self-energy functions $\overline{\varSigma}^{F,\rho}$.

\subsection{$1/\mathcal{N}$ expansion of the 2PI effective action}
\label{sec:2PI1N}

The Lagrangian \eq{lagrangiandesityforcomplexfields} is symmetric under global $U(1)$ transformations of the complex-valued fields $\psi_{\alpha}(x)$.
In the special case that all couplings $\lambda_{\alpha\beta}$ are equal%
\footnote{For the interactions considered here, \TwoEqs{N_species_V_int}{interactionpartofS}, the fermionic property of the interaction is implemented in the fields since $\hat\Psi_{a}(x)\hat\Psi_{a}(x) = \psi_{a}(x)\psi_{a}(x) = 0$; therefore, it is unnecessary to additionally require $\lambda_{\alpha\alpha} = 0$, and all couplings $\lambda_{\alpha\beta}$ can indeed be chosen to be equal.}, the Lagrangian has an additional $O(\mathcal{N})$, one has an additional $O(\mathcal{N})$ symmetry in the space of all hyperfine levels described by the multicomponent field $\psi_{\alpha}(x)$.
This symmetry can be used to derive an expansion of the 2PI part $\varGamma_{2}$ of the effective action in powers of the inverse number of hyperfine levels $\mathcal{N}$.

Consider the powers of $\mathcal{N}$ in the diagrams in \Fig{Gamma2LOAndNLO}.
\begin{figure}
  \begin{center}
    \includegraphics[width=.37\textwidth]{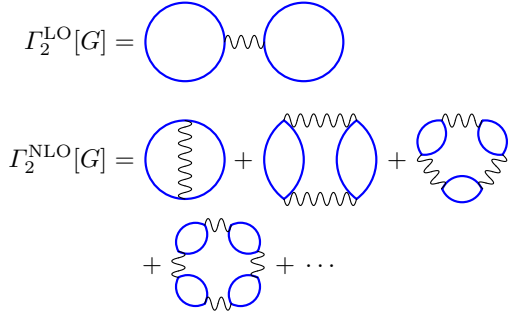}
  \end{center}
  \caption{ (Color online) 
    Diagrammatic representation of the leading order (LO) and next-to-leading order (NLO) contributions to $\varGamma_2$ in the $1/\mathcal{N}$ expansion.
    All statistical factors and relative minus signs are omitted.
  }
  \label{fig:Gamma2LOAndNLO}
\end{figure}
Each vertex contributes a factor of $1/\mathcal{N}$, see \Eq{lagrangiandesityforcomplexfields}.
Using the $O(\mathcal{N})$ symmetry, one can diagonalise the Green function for appropriate initial conditions, that is, spin balanced mode populations, such that $G_{\alpha\beta}=0$ for $\alpha\not=\beta$.
As a consequence, each loop contributes a factor of $\mathcal{N}$, and all diagrams with one new vertex appearing for each new loop contribute to $\varGamma_{2}$ at the same order.

In the following, we derive the leading-order (LO) and next-to-leading-order (NLO) terms of the 2PI $1/\mathcal{N}$ expansion,
\begin{equation}
  \varGamma_{2}[G]
  =\varGamma_{2}^{\text{LO}}[G]
   +\varGamma_{2}^{\text{NLO}}[G]
   +\dotsb\,.
  \label{eq:varGamma21rNsummation}
\end{equation}
The LO contribution is equivalent to one part of the Hartree-Fock-Bogoliubov (HFB) diagram,
\begin{equation}
  \varGamma_{2}^{\text{LO}}[G]
  = -\frac{\lambda_{\alpha\beta}}{8\mathcal{N}}
      \int_{x}
      G_{aa}(x,x) G_{bb}(x,x)\,.
\label{eq:definitionOfGamma2LO}
\end{equation}
As there are two sums over $\alpha,\beta \in\{1,\dotsc,\mathcal{N}\}$, this contribution is of the same order in $\mathcal{N}$ as the one-loop part of the action.
Hence, in the limit $\mathcal{N}\to\infty$, the dynamic equations contain less terms than in the HFB approximation, and the dynamics is entirely mean-field.

The NLO contribution reads
\begin{equation}
  \varGamma_{2}^{\text{NLO}}[G]
  =\frac{\myi}{2}\Tr{\ln\bigl[B(x,y;G)\bigr]}
\label{eq:definitionOfGamma2NLO}
\end{equation}
with
\begin{equation}
  B_{\alpha\beta}(x,y;G)
  = \delta_{\alpha\beta}\delta_{\mathcal{C}}(x-y)
     - \frac{\myi \lambda_{\alpha\gamma}}{2\mathcal{N}}
     \varPi_{\gamma\beta}(x,y)
\label{eq:definitionB}
\end{equation}
such that
\begin{align}
\begin{split}
  &\Tr{\ln\bigl[B(x,y;G)\bigr]}\\
  &=-\int_{x}
    \frac{\myi \lambda_{\alpha\beta}}{2\mathcal{N}}
    \varPi_{\beta\alpha}(x,x)
  \\
  &\quad\mbox{}
   -\frac{1}{2}
    \int_{xy}
      \frac{\myi \lambda_{\alpha\beta}}{2\mathcal{N}}
      \varPi_{\beta\gamma}(x,y)
      \frac{\myi \lambda_{\gamma\delta}}{2\mathcal{N}}
      \varPi_{\delta\alpha}(y,x)
  \\    
  &\quad\mbox{}
    - \dotsb\\
\end{split}
\end{align}
with $\varPi$ defined in Eq.~\eqref{eq:definitionOfPi}. 
In analogy to \Eq{varGamma21rNsummation}, the proper self-energy has LO and NLO contributions,
\begin{equation}
  \varSigma_{ab}(x,y;G)
  =\varSigma_{ab}^{\text{LO}}(x,y;G)
   +\varSigma_{ab}^{\text{NLO}}(x,y;G)
   +\dotsb\,,
\end{equation}
with
\begin{align}
  \varSigma^{\text{LO}}_{ab}(x,y;G)
  &= \delta_{ab}
     \frac{\myi \lambda_{\alpha\gamma}}{2\mathcal{N}}
     G_{cc}(x,x)
     \delta_{\mathcal{C}}(x-y)\,,
\\
  \varSigma^{\text{NLO}}_{ab}(x,y;G)
  &=-\frac{2\myi}{\mathcal{N}}
    \varLambda_{\alpha\beta}(x,y;G)
    G_{ab}(x,y) \,.
\end{align}
The NLO contribution can be understood as a scattering diagram with a resummed vertex
\begin{equation}
  \varLambda_{\alpha\beta}(x,y)
  = \Bigl(
       \delta_{\alpha\gamma}
       \delta_{\mathcal{C}}(x-y)
       + \myi I_{\alpha\gamma}(x,y;G)
    \Bigr)
    \frac{\lambda_{\gamma\beta}}{2} \,,
\label{eq:DefinitionOfLambda}
\end{equation}
which is defined through the integral equation
\begin{equation}
  I_{\alpha\beta}(x,y;G)
  = \frac{1}{\mathcal{N}}
      \int_{z}
      \varLambda_{\alpha\gamma}(x,z;G)
      \varPi_{\gamma\beta}(z,y)\,.
\label{eq:DefitionResummedBubbleChainI}
\end{equation}
One recovers the Hartree-Fock-Bogoliubov approximation by setting the resummed local interaction function $\varLambda$ equal to the bare coupling, $\varLambda_{\alpha\beta}(x,y)=\lambda_{\alpha\beta}\delta(x-y)/2$.

The self-energy up to NLO has in general both local and nonlocal contributions, the local one being equivalent to the HFB term,
\begin{equation}
  \varSigma_{ab}^{\text{HFB}}(x,y)
  = \varSigma_{ab}^{\text{LO}}(x,y)
    + \varSigma_{ab}^{\text{NLO}}(x,y)\bigr|_{\varLambda = g}\,.
\end{equation}
The nonlocal beyond-mean-field contribution to the self-energy is given by
\begin{equation}
  \overline\varSigma_{ab}(x,y;G)
  = I_{\alpha\gamma}(x,y;G)
    \frac{\lambda_{\gamma\beta}}{\mathcal{N}}
    G_{ab}(x,y)\,.
\end{equation}
The real functions $M_{ab}(x;G)$, $\overline\varSigma^{F}_{ab}(x,y;G)$, and $\overline\varSigma^{\rho}_{ab}(x,y;G)$ are all regular in $x_{0}$ and are obtained in terms of statistical and spectral functions as follows:
\begin{align}
\begin{split}
  M_{ab}(x)
  &= \delta_{ab}
     \Bigl(
       H^{\text{1B}}_{\alpha\beta}(\vector{x})
       - \frac{\lambda_{\alpha\gamma}}{2\mathcal{N}}
         F_{cc}(x,x)
      \Bigr)\\
  &\quad\mbox{}
       + \frac{\lambda_{\alpha\beta}}{\mathcal{N}}
         F_{ab}(x,x)\,,
\end{split}\\
\begin{split}
  \overline\varSigma^{F}_{ab}(x,y)
  &=\frac{\lambda_{\gamma\beta}}{\mathcal{N}}
    \Bigl(
      I^{F}_{\alpha\gamma}(x,y)
      F_{ab}(x,y)\\
  &\hspace{4em}\mbox{}
      - \frac{1}{4}
        I^{\rho}_{\alpha\gamma}(x,y)
        \rho_{ab}(x,y)
    \Bigr)\,, 
\end{split}\\
\begin{split}
  \overline\varSigma^{\rho}_{ab}(x,y)
  &= \frac{\lambda_{\gamma\beta}}{\mathcal{N}}
     \Bigl(
       I^{F}_{\alpha\gamma}(x,y)
       \rho_{ab}(x,y)\\
  &\hspace{4em}\mbox{}  
      + I^{\rho}_{\alpha\gamma}(x,y)
        F_{ab}(x,y)
     \Bigr)\,,
\end{split}
\end{align}
where
\begin{align}
\begin{split}
  I^{F}_{\alpha\beta}(x,y)
  &= \frac{\lambda_{\alpha\gamma}}{2\mathcal{N}}
     \varPi^F_{\gamma\beta}(x,y)\\
  &\quad\mbox{}
    + \frac{\lambda_{\delta\gamma}}{2\mathcal{N}}
      \biggl(
        \int^{x_{0}}_{0}\text{d}z\,
        I^{\rho}_{\alpha\delta}(x,z)
        \varPi^{F}_{\gamma\beta}(z,y)\\
  &\hspace{4.5em}\mbox{}
        - \int^{y_{0}}_{0}\text{d}z\,
          I^{F}_{\alpha\delta}(x,z)
          \varPi^\rho_{\gamma\beta}(z,y)
      \biggr)\,,
\end{split}\\
\begin{split}
  I^{\rho}_{\alpha\beta}(x,y)
  &= \frac{\lambda_{\alpha\gamma}}{2\mathcal{N}}
     \varPi^{\rho}_{\gamma\beta}(x,y)\\
  &\quad\mbox{}
     + \frac{\lambda_{\delta\gamma}}{2\mathcal{N}}
       \int^{x_{0}}_{y_{0}}\text{d}z\,
       I^{\rho}_{\alpha\delta}(x,z)
       \varPi^{\rho}_{\gamma\beta}(z,y)\,,
\end{split}
\end{align}
with $\varPi^F$ and $\varPi^\rho$ defined in Eqs.~\eqref{eq:definitionOfPiF} and \eqref{eq:definitionOfPiRho}, respectively.
As shown in \App{ConsLaws}, the dynamic equations derived from the 2PI effective action fulfill crucial conservation laws including those for the total particle number and the total energy.

In the next section, we focus on a homogeneous gas in a box with periodic boundary conditions, in which case the equations of motion in momentum space and in NLO $1/\mathcal{N}$ approximation are given by
\begin{align}
  &\Bigl(
    \myi\tau_{ac}\del_{x_{0}}
    -M_{ac}(x_{0};\vector{p})
  \Bigr)
  F_{cb}(x_{0},y_{0};\vector{p}) \nonumber\\
  &=\int^{x_{0}}_{t_{0}}\mathrm{d}z_{0}\,
    \overline\varSigma^{\rho}_{ac}(x_{0},z_{0};\vector{p})
    F_{cb}(z_{0},y_{0};\vector{p}) \nonumber\\
  &\quad\mbox{}
    -\int^{y_{0}}_{t_{0}}\mathrm{d}z_{0}\,
     \overline\varSigma^{F}_{ac}(x_{0},z_{0};\vector{p})
     \rho_{cb}(z_{0},y_{0};\vector{p})\,,
  \label{eq:EOFForFInMomentumSpace}
\\
  &\Bigl(
    \myi\tau_{ac}\del_{x_{0}}
    -M_{ac}(x_{0};\vector{p})
  \Bigr)
  \rho_{cb}(x_{0},y_{0};\vector{p}) \nonumber\\
  &=\int^{x_{0}}_{y_{0}}\mathrm{d}z_{0}\,
    \overline\varSigma^{\rho}_{ac}(x_{0},z_{0};\vector{p})
    \rho_{cb}(z_{0},y_{0};\vector{p})
 \label{eq:EOFForRhoInMomentumSpace}
\end{align}
with
\begin{align}
  M_{ab}(x_{0};\vector{p})
  &= \delta_{ab}
     \left(
       \frac{\vector{p}^{2}}{\n m}
       - \frac{\lambda_{\alpha\delta}}{2\mathcal{N}}
         \int_{\vector{k}}
         F_{dd}(x_{0},x_{0};\vector{k})
      \right) \nonumber\\
  &\quad\mbox{}
      + \frac{\lambda_{\alpha\beta}}{\mathcal{N}}
         \int_{\vector{k}}
         F_{ab}(x_{0},x_{0};\vector{k})\,,
\end{align}
and
\begin{align}
  &\overline\varSigma^{F}_{ab}(x_{0},y_{0};\vector{p}) \nonumber\\
  &=\frac{\lambda_{\gamma\beta}}{\mathcal{N}}
    \int_{\vector{k}}
    \Bigl(
      I^{F}_{\alpha\gamma}(x_{0},y_{0};\vector{p}-\vector{k})
      F^{}_{ab}(x_{0},y_{0};\vector{k}) \nonumber\\
  &\hspace{4.5em}\mbox{}
     - \frac{1}{4}
       I^{\rho}_{\alpha\gamma}(x_{0},y_{0};\vector{p}-\vector{k})
       \rho^{}_{ab}(x_{0},y_{0};\vector{k})
    \Bigr)\,,
\label{eq:SigmaFInMomentumSpace}\\
  &\overline\varSigma^{\rho}_{ab}(x_{0},y_{0};\vector{p}) \nonumber\\
  &= \frac{\lambda_{\gamma\beta}}{\mathcal{N}}
     \int_{\vector{k}}
     \Bigl(
       I^{F}_{\alpha\gamma}(x_{0},y_{0};\vector{p}-\vector{k})
      \rho^{}_{ab}(x_{0},y_{0};\vector{k}) \nonumber\\
  &\hspace{4.5em}\mbox{}
      + I^{\rho}_{\alpha\gamma}(x_{0},y_{0};\vector{p}-\vector{k})
        F^{}_{ab}(x_{0},y_{0};\vector{k}) 
     \Bigr)\,.
\label{eq:SigmaRhoInMomentumSpace}
\end{align}
Here, the functions $I^{F}_{}$ and $I^{\rho}_{}$ are given by
\begin{align}
  &I^{F}_{\alpha\beta}(x_{0},y_{0};\vector{p}) \nonumber\\
  &= \frac{\lambda_{\alpha\gamma}}{2\mathcal{N}}
     \varPi^{F}_{\gamma\beta}(x_{0},y_{0};\vector{p}) \nonumber\\
  &\quad\mbox{}
    + \frac{\lambda_{\delta\varepsilon}}{2\mathcal{N}}
      \biggl(
        \int^{x_{0}}_{t_{0}}\text{d}z_{0}\,
            I^{\rho}_{\alpha\delta}(x_{0},z_{0};\vector{p})
            \varPi^{F}_{\varepsilon\beta}(z_{0},y_{0};\vector{p}) \nonumber\\
  &\hspace{3em}\mbox{}
        - \int^{y_{0}}_{t_{0}}\text{d}z_{0}\,
          I^{F}_{\alpha\delta}(x_{0},z_{0};\vector{p})
          \varPi^{\rho}_{\varepsilon\beta}(z_{0},y_{0};\vector{p})
    \biggr)\,,
\label{eq:IFInMomentumSpace}\\
  &I^{\rho}_{\alpha\beta}(x_{0},y_{0};\vector{p}) \nonumber\\
  &= \frac{\lambda_{\alpha\gamma}}{2\mathcal{N}}
      \varPi^{\rho}_{\gamma\beta}(x_{0},y_{0};\vector{p}) \nonumber\\
  &\quad\mbox{}
     + \frac{\lambda_{\delta\varepsilon}}{2\mathcal{N}}
       \int^{x_{0}}_{y_{0}}\text{d}z_{0}\,
        I^{\rho}_{\alpha\delta}(x_{0},z_{0};\vector{p})
       \varPi^{\rho}_{\varepsilon\beta}(z_{0},y_{0};\vector{p})\,,
\label{eq:IRhoInMomentumSpace}
\end{align}
with
\begin{align}
  &\varPi^{F}_{\alpha\beta}(x_{0},y_{0};\vector{p}) \nonumber\\
  &= \int_{\vector{k}}
     \Bigl(
       F_{(\alpha,i)(\beta,j)}(x_{0},y_{0};\vector{p}-\vector{k})
       F_{(\beta,j)(\alpha,i)}(y_{0},x_{0};\vector{k}) \nonumber\\
  &\quad\mbox{}
       + \frac{1}{4}
         \rho_{(\alpha,i)(\beta,j)}(x_{0},y_{0};\vector{p}-\vector{k})
         \rho_{(\beta,j)(\alpha,i)}(y_{0},x_{0};\vector{k})
     \Bigr)\,,
\label{eq:PiFInMomentumSpace}\\
  &\varPi^{\rho}_{\alpha\beta}(x_{0},y_{0};\vector{p}) \nonumber\\
  &= \int_{\vector{k}}
     \Bigl(
       \rho_{(\alpha,i)(\beta,j)}(x_{0},y_{0};\vector{p}-\vector{k})
       F_{(\beta,j)(\alpha,i)}(y_{0},x_{0};\vector{k}) \nonumber\\
  &\quad\mbox{}
       - F_{(\alpha,i)(\beta,j)}(x_{0},y_{0};\vector{p}-\vector{k})
         \rho_{(\beta,j)(\alpha,i)}(y_{0},x_{0};\vector{k})
     \Bigr)\,,
\label{eq:PiRhoInMomentumSpace}
\end{align}
where $\int_{\vector{k}}=(2\pi)^{-d}\int\text{d}^{d}k$.
In one spatial dimension, these equations are conveniently written in dimensionless variables.
This is achieved by defining
  $\tilde t = n_\text{1D}^2 t/m$,
  $\tilde{\vector{p}} = \vector{p}/n_\text{1D}$,
  $\gamma_{\alpha\beta} = m g_{\alpha\beta}/n_\text{1D}$,
  $\tilde M = m M/n_\text{1D}^2$,
  $\tilde\varSigma = m^2 \varSigma/n_\text{1D}^4$ and
  $\tilde I = m I/n_\text{1D}^2$,
where the tilde denotes the rescaled quantities, $\gamma_{\alpha\beta}$ is the dimensionless coupling constant and $n_\text{1D}$ denotes the line density.

\section{Nonequilibrium time evolution of a one-dimensional Fermi gas}
\label{sec:Results1D}

In this section, we apply the dynamic equations derived from the 2PI effective action, in next-to-leading order (NLO) of the $1/\mathcal{N}$ expansion, to the case of a homogeneous ultracold Fermi gas with twofold hyperfine degeneracy (denoted as $\uparrow$ and $\downarrow$) in one spatial dimension.
We study the system in a finite-size box with periodic boundary conditions. 
The Fermi gas is initially assumed to be noninteracting and prepared far from equilibrium, characterised by a nonequilibrium single-particle momentum distribution.
The equations of motion, \TwoEqs{EOFForFInMomentumSpace}{EOFForRhoInMomentumSpace}, are solved in momentum space.
We assume the interactions to be switched on at the initial time and investigate the long-time evolution of the interacting gas towards equilibrium.

The homogeneous one-dimensional gas is taken to have a line density $n_\text{1D}$, and the constituents have mass $m$.
To identify the relevant combination of parameters, it is convenient to rewrite the equations of motion in dimensionless variables, and to introduce the dimensionless coupling constant $\gamma_{\alpha\beta} = m g_{\alpha\beta}/n_\text{1D}$.

\subsection{Initial conditions}

With initial values for the spectral function $\rho(t_0,t_0;p)$ and the statistical propagator $F(t_0,t_0;p)$, Eqs.~\eqref{eq:EOFForFInMomentumSpace} and \eqref{eq:EOFForRhoInMomentumSpace} describe the time evolution of the two-time correlation functions including the momentum distribution
\begin{equation}
  n_{\alpha}(t,p)
   = \frac{1}{2}
   \big(
     1 - F^{}_{(\alpha,i)(\alpha,i)}(t,t;p)
   \big)\,.
  \label{eq:momentumDistribution}
\end{equation}
In the following, we will choose different initial momentum distributions $n_{\alpha}(t_{0},p)$ away from thermal equilibrium.
Furthermore, we choose the initial coherence between different spins as well as the initial pair correlation function to vanish,
\begin{align}
  \langle
    \hat\Psi^{\dagger}_{\alpha}(\vector{x},t_0)
    \hat\Psi^{}_{\beta}(\vector{x},t_0)
  \rangle
  &= 0 \qquad\text{for $\alpha\neq\beta$,}
 \\
 \langle \hat\Psi^{}_{\alpha}(\vector{x},t_0) \hat\Psi^{}_{\beta}(\vector{x},t_0) \rangle
 &= 0 \,.
\end{align}
For $\alpha=\beta$, the equal-time pair correlation function always vanishes, which is in accordance with the conservation of the total particle number and a direct consequence of the equal-time property of the spectral function (see App.~\ref{app:2pGreenF}).
For $\alpha\neq\beta$, a nonzero initial pair correlation function would account for BCS-type pairs and imply a nonzero variance of the total particle number.
The above initial conditions require
\begin{align}
  F_{(\alpha,i_{a}){(\beta,i_{b})}}(t_{0}, t_{0}; p)
  &= 0 \qquad\text{for $i_{a}\neq i_{b}$.}
  \label{eq:F11=F22}
\end{align}

Combining Eqs.~\eqref{eq:momentumDistribution} and~\eqref{eq:F11=F22} yields the initial condition
\begin{align}
  F^{}_{(\alpha,1)(\alpha,1)}(t_0, t_0; p)
  &= F^{}_{(\alpha,2)(\alpha,2)}(t_0, t_0;p) \nonumber\\
  &= \frac{1}{2} - n_{\alpha}(t_0,p)\,.
\end{align}
The equal-time property of the spectral function, Eq.~\eqref{eq:equaltimeanticommutator}, requires
\begin{equation}
 \rho_{ab}(t_{0}, t_{0};p) = \myi\tau_{ab} \,.
\end{equation} 

For a homogeneous gas and a ($p \leftrightarrow -p$)-symmetric initial state, $F$ and $\rho$ are invariant under $p \rightarrow -p$ at all later times.

\subsection{Numerical method}

We numerically solve the equations of motion~\eqref{eq:EOFForFInMomentumSpace} and \eqref{eq:EOFForRhoInMomentumSpace} together with the nonperturbative integral equations for the self-energies~\eqref{eq:SigmaFInMomentumSpace} and \eqref{eq:SigmaRhoInMomentumSpace}.
Due to the memory integrals, computations are costly, and in computing the results shown below, we kept a finite memory kernel at longer times, checking that an increase in the memory time did not change the results.
We also carefully checked that a change in the size of the box and the number of the momentum modes does not lead to a significant change in the results presented in the following, except for finite-size (infrared-cutoff) effects taken into account explicitly.
For example, we chose the interaction strength sufficiently weak such that the occupation numbers of modes close to the cutoff defined by the grid size are small enough not to give rise to ultraviolet-cutoff-dependent effects.
For the results presented here, we discretised the momentum space on a lattice with $N_s = 128$ sites and periodic boundary conditions.
Thus, the momentum modes corresponding to the lattice Laplacian are $p_{j}/n_{\text{1D}} = 2N_s\sin(j\pi/N_s)/N$, where $j\in\{-N_s/2+1,-N_s/2+2,\dotsc,N_s/2\}$ and $N=\sum_{\alpha,j}n_\alpha(p_j)$ is the total particle number.

The diagonal time steps were implemented according to
\begin{equation}
\begin{split}
  F(t_{n+1}, t_{n+1}; p)
  &= F(t_{n}, t_{n}; p) \\
  &\quad\mbox{}
     + \Bigl(
       F(t_{n+1}, t_{n}; p)
       - F(t_{n}, t_{n}; p)
     \Bigr) \\
  &\quad\mbox{}
     + \Bigl(
         F(t_{n}, t_{n+1}; p)
         - F(t_{n}, t_{n}; p)
       \Bigr)
\end{split}
\end{equation}
in order to ensure that the equations of motion, which are implemented on discrete grid also in the time domain, satisfy the same global $U(1)$ symmetry as the continuous equations.
In this way, the total particle number is numerically exactly conserved.

\subsection{Results}

To begin with, we studied the time-evolution for two initial states with the same total particle number and total energy (and interaction strength $\gamma=4$), but different far-from-equilibrium momentum distributions.
The two initial momentum distributions are shown in Fig.~\ref{fig:RunAAndB}a.

Once the system evolves in time, multiple scattering events lead to a redistribution of momenta until the system reaches an equilibrated state.
For runs A and B, Fig.~\ref{fig:RunAAndB}b shows the momentum-mode occupation numbers as a function of time $t$ for six of the momentum modes.
We find that the equilibrated states at late times have the same momentum-mode occupation numbers.
Note that a conservative estimate of the recurrence time on the basis of the slowest oscillating discrete mode of the free gas is two orders of magnitude larger than the shown total evolution time.
The final ($t n_{\text{1D}}^2/m = 10$) momentum distributions are also shown in Fig.~\ref{fig:RunAAndB}a.
However, to see that the final states are absolutely identical, one has not only to look at the momentum-mode occupation numbers, but also to compare the nonlocal-in-time behaviour of the two-point functions $F_{ab}(x_0,y_0;p)$ and $\rho_{ab}(x_0,y_0;p)$ of the two runs for same center time coordinates $X\equiv(x_0+y_0)/2$ but different relative time coordinates $s\equiv x_0-y_0$.
Exemplarily, $F_{(\uparrow,1)(\uparrow,1)}(X,s;p)$ at late times ($n_{\text{1D}}^2 X/m = 10$) is depicted in Fig.~\ref{fig:RunAAndB}c for four of the momentum modes of runs A and B.
The data for the two runs lie on top of each other; thus, the final correlation functions of runs A and B are indeed identical.

\begin{figure}[t]
  \begin{center}
    \includegraphics[width=.37\textwidth]{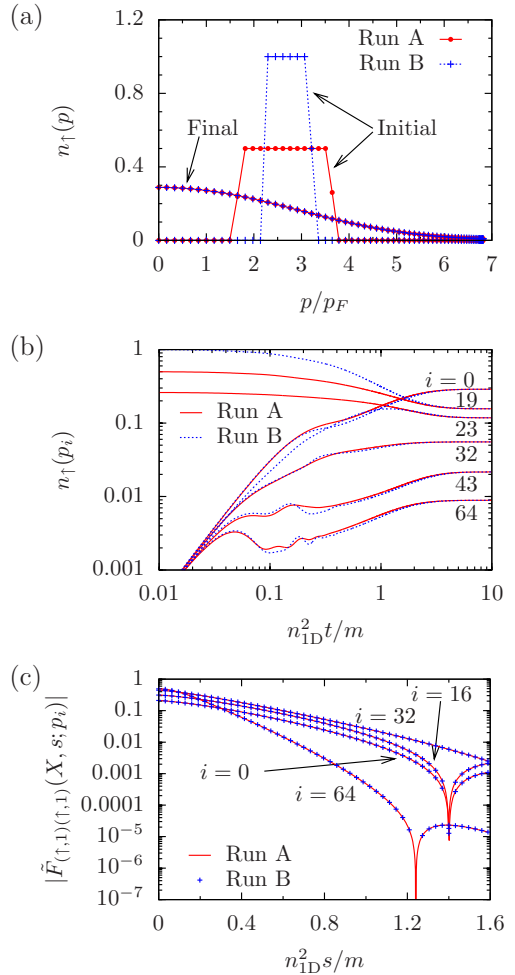}
  \end{center}
  \caption{ \label{fig:RunAAndB} (color online)
    Comparison between run A and run B.
    (a)
    Initial ($n_{\text{1D}}^2 t/m = 0$) and final ($n_{\text{1D}}^2 t/m = 10$) momentum-mode occupation numbers $n_\uparrow(t,p) = n_\uparrow(t,-p) = n_\downarrow(t,p)$.
    The initial states are set up such that both runs have the same total particle number and total energy.
    Therefore, the equilibrated states at late times have the same momentum distribution.
    (b)
    Numerically determined momentum-mode occupation numbers as a function of time $t$ for some of the momentum modes.
    Note that time and occupation numbers are shown on a logarithmic scale.
    Note that a conservative estimate of the recurrence time on the basis of the slowest oscillating discrete mode of the free gas is two orders of magnitude larger than the shown total evolution time.
    (c)
    Nonlocal-in-time behaviour of the envelope functions
      $\tilde{F}_{(\uparrow,1)(\uparrow,1)}(X,s;p)$
    of the statistical propagator
      $F_{(\uparrow,1)(\uparrow,1)}(X,s;p)
       = \tilde{F}_{(\uparrow,1)(\uparrow,1)}(X,s;p)
         \cos[p^2s/(\sqrt{2}m)]$
    at late center times ($n_{\text{1D}}^2 X/m =10$) for some of the momentum modes. The data lie on top of each other, which proofs that the final states of both runs are identical.
  }
\end{figure}

Furthermore, to observe that the final momentum distribution has the form of  a Fermi-Dirac distribution within the range of momenta considered,
it is convenient to look at the inverse slope function $\sigma = \ln[1/n_{\uparrow}-1]$ rather than the occupation numbers directly.
When substituting the inverse slope function for the exponent of the Fermi-Dirac distribution, i.\,e.,
\begin{equation}
  n_{\uparrow}(t, p) = \frac{1}{\exp\{ \sigma[\omega(p)] \}+1}\,,
\end{equation}
it is apparent that $\sigma$, as a function of mode energy $\omega(p)$, reduces to a straight line when the occupation number $n_{\uparrow}(t, p)$ is a Fermi-Dirac distribution.
Fig.~\ref{fig:RunBAnalysis}a shows the inverse slope function as a function of mode energy $\omega(p)=p^{2}/2$ for run B at various times $t$.
At late times, where the inverse slope function becomes a straight line, one can extract the inverse temperature $\beta$ and the chemical potential $\mu$ according to
  $\sigma(\omega) = \beta ( \omega - \mu )$.
\begin{figure}
  \begin{center}
    \includegraphics[width=.47\textwidth]{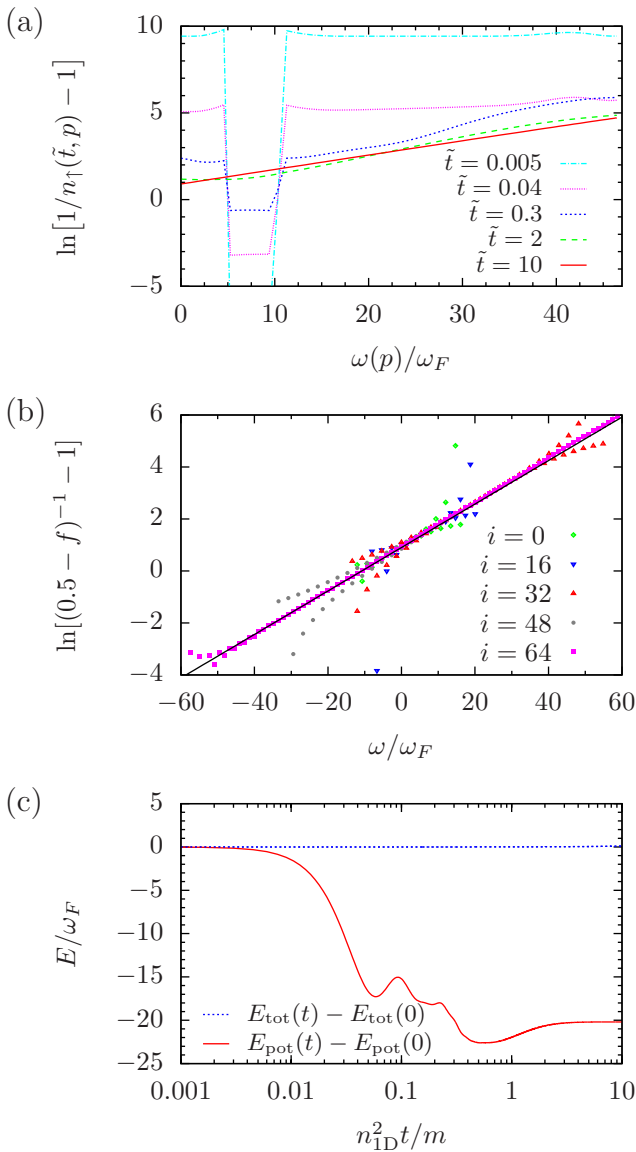}
  \end{center}
  \caption{ \label{fig:RunBAnalysis} (color online)
    Further analysis of run B.
    (a)
    Fermion distribution $n_{\uparrow}(t, p)$ as a function of mode energy $\omega(p)$ in units of the Fermi energy $\omega_F=\omega(p_F)$ at various times $\tilde t = n_{\text{1D}}^2 t/m$.
    Plotted is the inverse slope function $\sigma=\ln[1/n_{\uparrow}-1]$, which reduces to a straight line when $n_{\uparrow}(t,p)$ approaches a Fermi-Dirac distribution.
    (b)
    Fluctuation-dissipation relation at late times ($n_{\text{1D}}^2 t/m=10$) for several momentum modes $\tilde{p}_i$.
    The fraction $f$ is defined in Eq.~\eqref{eq:fractionF}.
    (c)
    Time evolution of the potential and total energy.
    The reduction of the initial potential energy is due to the built-up of additional kinetic energy.
}
\end{figure}

A thermal state, however, does not necessarily require that the single-particle momentum distribution coincides with a Fermi-Dirac distribution.
For the  state to be described as a (grand-)canonical ensemble, a necessary condition is the Callen-Welton fluctuation-dissipation relation \cite{Callen1951a,Jin2007a}
\begin{equation}
  F_{\alpha\alpha}(X_{0};\omega,{p})
   = -\myi[ 1/2-n_\mathrm{FD}(\omega-\mu) ] \rho_{\alpha\alpha}(X_{0};\omega,{p})
\label{eq:flucDisRel}
\end{equation}
which connects the statistical propagator $F$ and the spectral function $\rho$.
Here,
  $F_{\alpha\alpha}(X;\omega,p)
   = \int\text{d}s
       \exp(\myi\omega s)
       F_{\alpha\alpha}(X+s/2, X-s/2;p)$, 
  $F_{\alpha\alpha}(t,t';p)
   =\langle
      [\hat\Psi^\dagger_{\alpha}(t,p),
       \hat\Psi_{\alpha}(t',p)]_-
    \rangle/2$,
and similar for
  $\rho_{\alpha\alpha}(t,t';p)
   = \myi\langle
       [\hat\Psi^\dagger_{\alpha}(t,p),
        \hat\Psi_{\alpha}(t',p)]_+
     \rangle$.
We emphasise that Eq.~\eqref{eq:flucDisRel} is valid for a grand canonical state irrespective of whether the system is interacting, i.\,e., whether single-particle modes are eigenmodes of the Hamiltonian, or not.

Figure~\ref{fig:RunBAnalysis}b shows the emergence of the fluctuation-dissipation relation at late times for run~B,
where we introduced the fraction
\begin{equation}
  f = \myi F_{\uparrow\uparrow}(X_{0};\omega,p)/\rho_{\uparrow\uparrow}(X_{0};\omega,p)
  \label{eq:fractionF}
\end{equation}
 as a function of the frequency $\omega$.
 $f$ is shown in a region around the peaks of the statistical and spectral functions where the argument of the logarithm $\ln[(1/2-f)^{-1}-1]$ is positive.
Outside this region it oscillates around zero due to finite evolution time after the quench.
Propagating the dynamic equations further reduces these oscillations.
Hence, according to the fluctuation-dissipation theorem, the system is approximately thermalised over the depicted range of energies.

The redistribution of the initial kinetic and potential energies during the time evolution is shown in Fig.~\ref{fig:RunBAnalysis}c.
Since the total energy is calculated by summing the numerically determined kinetic and potential energies, this plot also highlights the numerical accuracy of the conservation of the total energy.
We have explicitly checked that extending the size of the memory kernel further does not change our results.

In conlusion, the chosen initial conditions for runs A and B allow the single-particle momentum distributions to thermally equilibrate over the considered range of momenta.
We find that the final state is determined by the values of the conserved quantities in the initial state.
All other information about the details of the initial state is lost during the evolution.

Next, we investigate a few thermodynamic properties of the equilibrated interacting Fermi gas.
For this purpose, we performed additional runs with different initial energies but the same total particle number and interaction strength as in runs A and B.
Exemplarily, the initial momentum distributions of two of these runs are shown in Fig.~\ref{fig:RunCAndD}a together with the one from run B.
Even though all of the runs virtually reach a stationary state (Fig.~\ref{fig:RunCAndD}b) within the times that are numerically accessible before the time discretisation leads to a break-down of the energy conservation, the runs with lower initial energies do not fully settle to a Fermi-Dirac distribution.
As depicted in Fig.~\ref{fig:RunCAndD}c, the lower momentum modes reach a Fermi-Dirac distribution, but the higher momentum modes still show an excess population.
Keeping the total particle number constant, a further reduction of the population in the higher momentum modes will not significantly alter the population in the lower momentum modes since the former are already populated much less than the latter.
Therefore, we can extract temperatures and chemical potentials from a fit of the lowest 14 momentum modes to a Fermi-Dirac distribution as it is shown in Fig.~\ref{fig:RunCAndD}c.
\begin{figure}
  \begin{center}
    \includegraphics[width=.47\textwidth]{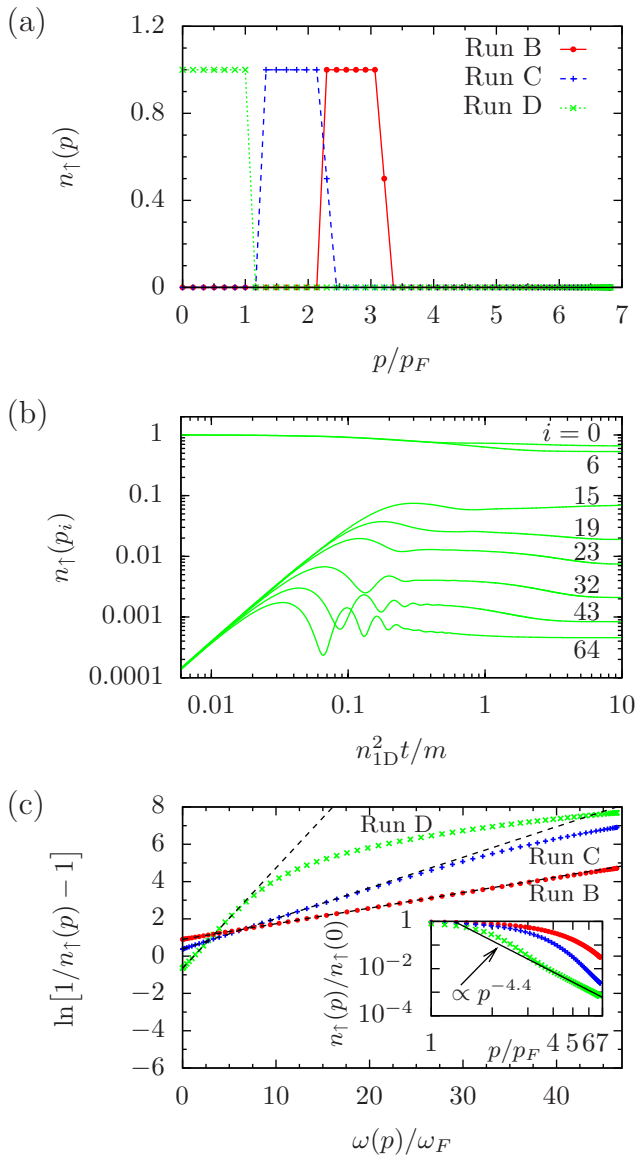}
  \end{center}
  \caption{ \label{fig:RunCAndD} (color online)
    (a)
    Initial momentum distributions of runs B, C, and D.
    The runs have the same total particle number but different total energies.
    (b)
    Time evolution of the occupation numbers for some of the momentum modes of run D.
    Similar to all other performed runs, a stationary state is reached at late times.
    (c)
    Inverse slope functions for runs B, C, and D at late times.
    The black dashed lines are fits to the lowest 14 momentum modes from which the temperatures and chemical potentials are extracted.
    The inset depicts the power-law tail in the momentum distributions for low energies.
    }
\end{figure}

The so found temperature dependence of the late-time kinetic energy $E_{\text{kin}}^{\text{(eq)}}$, the heat capacity $C_V=k_B \del E_{\text{tot}}/\del\beta^{-1}$ at constant volume, and the chemical potential $\mu$ are shown in Fig.~\ref{fig:TempVsKinEnergyAndHeatCapacityAndChemicalPotential}.
At high temperatures, the quantities of the interacting gas converge towards the results for an ideal Fermi gas.
However, at low temperatures, they significantly deviate from those of an ideal gas due to the finite coupling constant as shown in the insets.
Note also that the results are sensitive to the finite size of our system.
Therefore, also the results for the ideal Fermi gas in discrete momentum space differ from those of an ideal Fermi gas in the thermodynamic limit, where the momentum space is continuous.
\begin{figure}
  \begin{center}
    \includegraphics[width=.47\textwidth]{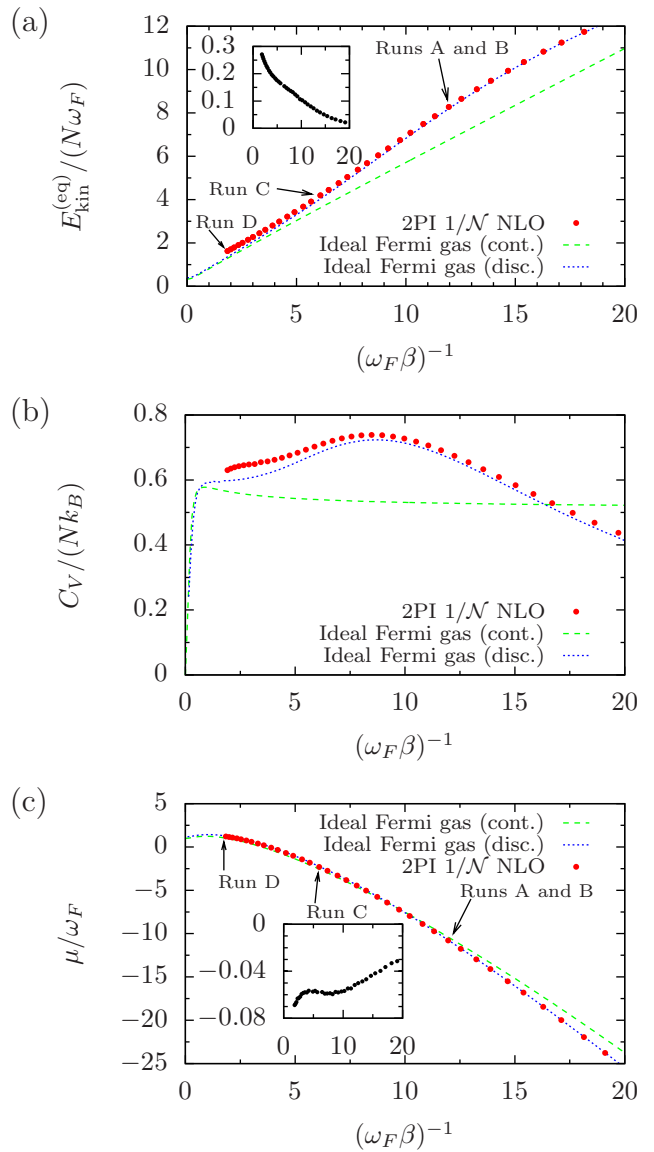}
  \end{center}
  \caption{  \label{fig:TempVsKinEnergyAndHeatCapacityAndChemicalPotential} (color online)
    Temperature dependence of
      (a) the mean kinetic energy per particle,
      (b) the heat capacity per particle, and
      (c) the chemical potential.
    The results for the interacting Fermi gas from the 2PI $1/\mathcal{N}$ NLO runs are shown as red dots.
    For comparison, the exact results for the ideal Fermi gas are shown both for a continous and a discrete momentum space.
    The differences between the interacting Fermi gas and the ideal Fermi gas in discrete momentum space are shown in the insets.}
\end{figure}

Finally, we investigate the overpopulation of the occupation numbers for runs C and D at high momenta as compared to a Fermi-Dirac distribution.
As depicted in the inset of Fig.~\ref{fig:RunCAndD}c, the tail in the momentum distribution is characterised by a power-law $n_{\uparrow}(p)\propto p^{-\kappa}$ with $\kappa\simeq 4.4$.
This exponent does not change when diagrams of order $\lambda^3$ are included in the effective action, see \Fig{LL}.
Note, however, that it is crucial to go beyond the LO $1/\mathcal{N}$ and the HFB approximations of the effective action since both of them exclude multiple scattering events and thereby leave all momentum-mode occupation numbers unchagend during the time evolution, cf.\ \Eq{SolHFBEqn4F} and the ensuing discussion.
\begin{figure}[tb]
  \begin{center}
    \includegraphics[width=0.45\textwidth]{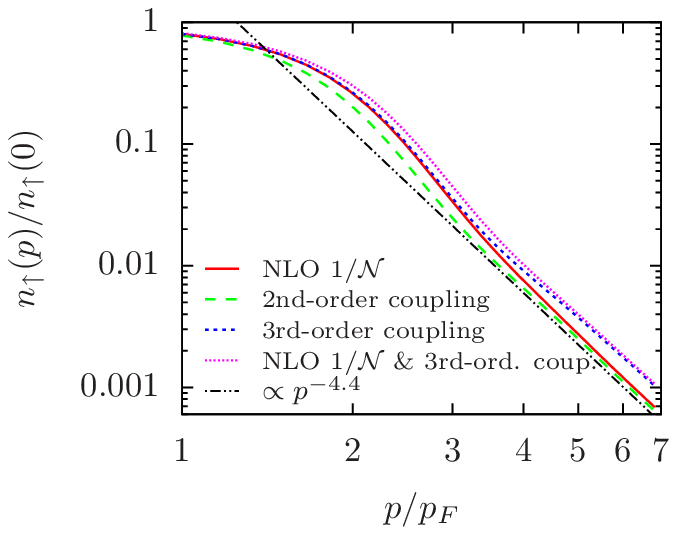}
  \end{center}
  \caption{  \label{fig:LL} (color online)
  Momentum distributions at $n_{\text{1D}}^2 t/m=6.4$ for runs with the same initial state as run D but in different approximations of the effective action.
}
\end{figure}
\begin{figure}[tb]
  \begin{center}
    \includegraphics[width=0.47\textwidth]{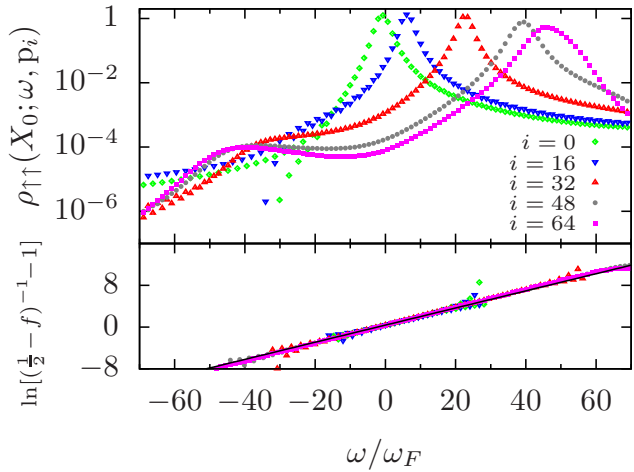}
  \end{center}
  \caption{  \label{fig:FDT} (color online)
  Top: Spectral functions as a function of frequency at late time $X_{0}=18.9\,n_{\text{1D}}^{-2}m$ in runs C for five of the momentum modes $\mathrm{p}_{i}$.
  Bottom: Inverse-slope function of fractions $f$ of the statistical $F$ divided by the spectral function $\rho$ at $X_{0}=18.9\,n_{\text{1D}}^{-2}m$, for the same five momentum modes.
  Black lines indicate Fermi-Dirac distributions with $\beta$ and $\mu$ as in \Fig{RunCAndD}c.
}
\end{figure}
\begin{figure}[tb]
  \begin{center}
    \includegraphics[width=0.47\textwidth]{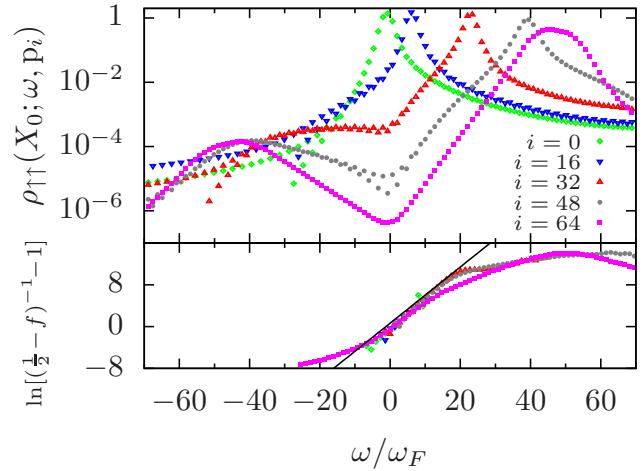}
  \end{center}
  \caption{  \label{fig:FDT2} (color online)
  Same as \Fig{FDT} but for run D.
  In run D, the system does not thermalise.
  Note the exponential decay of the spectral functions away from the peaks.
}
\end{figure}

As pointed out before, this power-law tail suggests but does not prove that the Fermi gas approaches a non-thermal state.
Therefore, we have a closer look at the fluctuation-dissipation relation, Eq.~\eqref{eq:flucDisRel}.
The lower panel of \Fig{FDT} shows the inverse-slope function $\ln[(1/2-f)^{-1}-1]$ of the fraction $f$ defined in Eq.~\eqref{eq:fractionF} for run C, the lower panel of  \Fig{FDT2} for run D.
$f$ is shown in a region where the argument of the logarithm is positive -- outside this region, it oscillates around zero as a result of the finite total evolution time after the quench.

In run C, the inverse-slope function is a straight line over the region of relevant $\omega$ and therefore corresponds to a Fermi-Dirac function.
As in run B, the system is thermalised over the depicted range of energies, in spite of the signs of a power-law tail in run C.
This can be understood by considering the spectral function in the upper panel of \Fig{FDT}: 
the second peak at negative frequencies picks up extra contributions from the Fermi sea thereby causing the power-law overpopulation at high momenta.
Although the area under the negative-$\omega$ peak is reduced by a factor of $\sim 10^{-4}$, it is multiplied by 1 on the filled-sea side of the Fermi-Dirac function while the positive-$\omega$ peak multiplies the exponentially suppressed tail of the Fermi-Dirac function.
Thus, one may preconclude from run C that the system thermalises to a grand-canonical ensemble, with the eigenmodes of the strongly interacting system at low temperatures being superpositions of particles and holes.
This contains reminiscence of the Bogoliubov depletion at zero temperature that gives, for a Fermi gas in the BCS theory, a $p^{-4}$ power-law tail of the single-particle momentum distribution.

However, run D performed at even lower energy shows that the system does in general not thermalise to a grand-canonical ensemble: even though the momentum overpopulation is again largely produced by the contributions from the Fermi sea, see \Fig{FDT2} (upper panel), also the fraction $f$ shows a power-law tail $\sim {p}^{-9}$ violating the fluctuation-dissipation theorem, see \Fig{FDT2} (lower panel).
Despite this, the equilibrated momentum distribution $\propto {p}^{-4.4}$ at low total energies is still mainly due to the second peak in the spectral function while the contribution from the nonthermal power-law tail of $f$ is suppressed by another 4.5 powers of $p$.

\section{Conclusions}
\label{sec:Concl}

We have summarised the description of far-from-equilibrium dynamics of ultracold Fermi gases in terms of Kadanoff-Baym equations for two-point many-body Green functions derived from the two-particle-irreducible (2PI) effective action in nonperturbative approximation.
The approach allows to handle both mean-field and beyond mean-field approximations on the same footings and to derive approximations far beyond mean-field and perturbative kinetic approaches in an elegant way.
Obtaining approximations on the level of the 2PI effective action ensures the conservation of energy irrespective of the chosen truncation as well as other vital quantities as the total particle number.
This forms a precondition for the applicability  of the approach for long-term evolution and equilibration.
Implicitly contained higher-order correlations render the dynamic equations nonlocal in time, causing a non-Markovian scattering integral.

Beyond the mean-field truncation which excludes effects of scattering between quasiparticles, we consider the nonperturbative expansion in inverse powers of the number of internal or spin degrees of freedom $\mathcal{N}$.
This is possible when the external potential and collisional interactions show an $\mathcal{N}$-fold spin degeneracy, i.\,e., if the Hamiltonian is symmetric under the orthogonal group $O(\mathcal{N})$.
We have considered a system of Fermions interacting through $s$-wave collisions between different spin components. 

As a specific example, we have studied the long-time evolution of a homogeneous, one-dimensional, twofold spin-degenerate Fermi gas with an initial far-from-equilibrium momentum distribution.
We have numerically solved the dynamical equations in next-to-leading order in the $1/\mathcal{N}$ expansion of the 2PI effective action.
Results presented for this extend upon the work presented in \cite{Kronenwett2010a}.
They give that, within the truncation considered, the one-dimensional gas dephases and equilibrates following an interaction quench.
For sufficiently high total energies, the system is found to equilibrate to a state with thermodynamic properties like chemical potential and specific heat given by those of a thermal ideal Fermi gas.
In contrast, close to the Fermi energy the equilibration leads to non-thermal power-law momentum distributions pointing to the appearance of many-body quasiparticle modes.
Furthermore, we found a violation of the fluctuation-dissipation relation for a grand-canonical ensemble which gives a strong indication of a nonthermal equilibrium state.
A future task is to extend our study to account for the limit of low energies near zero temperature, in particular for features of the Tomonaga-Luttinger low-energy fixed point, in order to clarify the transition regime between high and low energies.

\begin{acknowledgement}
The authors would like to thank C.~Bodet, M.~Holland, S.~Jochim, S.~Kehrein, V.~Meden, B.~Nowak, J.\,M.~Pawlowski, A.\,M.~Rey, and D.~Sexty for inspiring and useful discussions, and JILA and the University of Colorado for their hospitality.
They acknowledge the support by the Deutsche Forschungsgemeinschaft, as well as the support of the Alliance Program of the Helmholtz Association (HA216/EMMI).
M.\,K.\ thanks the Heidelberg Graduate School for Fundamental Physics (HGSFP) and the German Academic Exchange Service (DAAD) for financial support during his stay at JILA.
\end{acknowledgement}

\begin{appendix}

\section{Gra\ss mann variables}
\label{app:grassmannVariables}

In this appendix, we outline properties of Gra\ss mann variables that are needed in the context of our discussion of nonrelativistic fermionic path integrals and not commonly found in textbooks.
For a more detailed discussion of Gra\ss mann variables in the context of path integrals, we refer, e.\,g., to Ref.~\cite{ZinnJustin2004a}. 

A set of complex Gra\ss mann variables $\theta_i$, $i\in\{1,\dotsc,n\}$, satisfies
\begin{align}
  \theta_i \theta_j + \theta_j \theta_i &= 0\,,
  & i,j &\in\{1,2,\dotsc,n\} \,.
\label{eq:anticommutationGrassmannVariables}
\end{align}
This implies
\begin{equation}
  \theta_i \theta_i = 0 \,,
\end{equation}
with no summation over $i$.

For any pair of Gra\ss mann variables $\theta$ and $\vartheta$, complex conjugation (denoted by an asterisk) is defined by
\begin{equation}
  (\theta\vartheta)^* = \vartheta^* \theta^*\,,
\end{equation}
which ensures the reality condition $(\theta\theta^* )^* = \theta\theta^*$.

The real and imaginary parts of a complex Gra\ss mann variable $\theta$ are defined as
\begin{align}
  \frac{1}{\sqrt{\n}} \theta_1
  &=     \text{Re}[\theta]
  \equiv \frac{1}{2}\Bigl( \theta + \theta^* \Bigr)\,, \\
  \frac{1}{\sqrt{\n}} \theta_2
  &=     \text{Im}[\theta]
  \equiv \frac{1}{2\myi}\Bigl( \theta - \theta^* \Bigr)\,.
\end{align}
From this definition of $\theta_1$ and $\theta_2$, and property~\eqref{eq:anticommutationGrassmannVariables} of $\theta$ and $\theta^*$ follows immediately
\begin{align}
  \theta_1 \theta_1
    &= \theta_2 \theta_2
    =  0 \,, \\
  \myi \theta_1 \theta_2
    &= -\myi \theta_2 \theta_1
    =  \theta^* \theta\,,
\end{align}
which implies that $\myi \theta_1 \theta_2$ rather than $\theta_1 \theta_2$ is real.
This motivates to introduce the bared notation,
\begin{align}
  {\overline\theta}_1 &= -\myi \theta_2 \,,
  &{\overline\theta}_2 &= \myi \theta_1 \,,
\end{align}
so that
\begin{align}
  \theta_i {\overline\theta}_j
    &= -{\overline\theta}_j \theta_i
    \in\mathbb{R}\,,
  &i, j &\in \{1, 2\}\,,
\end{align}
and
\begin{equation}
  \theta^* \theta
  = \frac{1}{\n}
    \Bigl(
      {\overline\theta}_1 \theta_1
      + {\overline\theta}_2 \theta_2
    \Bigr) \,.
\end{equation}

\section{Two-point Green functions}
\label{app:2pGreenF}

In this appendix, we would like to summarise symmetry properties of the two-point Green function $G$ used in our derivations, as well as the statistical and spectral components of the two point function, $F$ and $\rho$, respectively.

The two-point Green function $G$ is defined as the time-ordered expectation value of two fields at two points in space-time,
\begin{equation}
  G_{ab}(x,y)
  =\langle
     \mathcal{T}_{\mathcal{C}}
     \hat\Psi_{a}(x)
     \hat{\overline\Psi}_{b}(y)
  \rangle\,,
  \label{eq:defGreensFunction}
\end{equation}
where $\mathcal{T}_{\mathcal{C}}$ denotes time-ordering along the closed time path $\mathcal{C}$.
Since there are no fermionic field expectation values, $G$ is automatically connected.
For Gra\ss mann fields, the time-ordering is defined as
\begin{equation}
\begin{split}
  &\mathcal{T}_{\mathcal{C}}\hat\Psi_{a}(x)\hat{\overline\Psi}_{b}(y)\\
  &=\begin{cases}
     \hat\Psi_{a}(x) \hat{\overline\Psi}_{b}(y)
     &\text{ if $\sgnC{x_{0}-y_{0}}=1$}\\
     - \hat{\overline\Psi}_{b}(y) \hat\Psi_{a}(x)
       &\text{ if $\sgnC{x_{0}-y_{0}}=-1$}\,,
   \end{cases}
\end{split}
\end{equation}
where $\sgnC{x_{0}-y_{0}}$ denotes the sign function along the time path $\mathcal{C}$ and evaluates to $1$ ($-1$) if $x_{0}$ is posterior (prior) to $y_{0}$.

The path integral formulation naturally accounts for time-ordering in expectation values of operator products.
At equal times, however, the time-ordering is ill-defined.
With the above definition the two-point Green function is singular at $x_{0}=y_{0}$.
For numerical implementations, it is convenient to make the singularity explicit and to decompose $G$ into its non-singular spectral and statistical components.
The spectral function $\rho$ contains information about the spectrum of the theory, i.\,e., the energies and decay times of the states, and the statistical propagator $F$ accounts for the respective occupation numbers.
They are defined as
\begin{align}
  F_{ab}(x,y)
  &\equiv \frac{1}{2}\langle\bigl[
                              \hat\Psi_{a}(x), \hat{\overline\Psi}_{b}(y)
                            \bigr]_{-}\rangle\,,
\label{eq:definitionOfF}\\
  \rho_{ab}(x,y)
  &\equiv \myi\langle\bigl[
                     \hat\Psi_{a}(x), \hat{\overline\Psi}_{b}(y)
                   \bigr]_{+}\rangle\,,
\label{eq:definitionOfRho}
\end{align}
where $[\cdot,\cdot]_{-}$ denotes the commutator and $[\cdot,\cdot]_{+}$ the anticommutator.
The decomposition identity then reads
\begin{equation}
  G_{ab}(x,y)
  = F_{ab}(x,y) - \frac{\myi}{2}\rho_{ab}(x,y)\sgnC{x_{0}-y_{0}} \,.
  \label{eq:decomposiontID4Gv2}
\end{equation}
The decomposition of $G$ into $F$ and $\rho$ has the advantage that the time ordering, and thus the discontinuity at $x_0 = y_0$, is accounted for explicitly by the sign function.

We close this appendix with some useful symmetry properties of the two-point functions:
\begin{subequations}
\begin{align}
  G_{ab}(x,y)    &= \tau_{bc} G_{cd}(y,x) \tau_{da}\!
                  =\! (-1)^{i_a + i_b} G_{\bar b\bar a}(y,x)\,,\! \\
  F_{ab}(x,y)    &= \tau_{bc} F_{cd}(y,x) \tau_{da}
                  =\! (-1)^{i_a + i_b} F_{\bar b\bar a}(y,x)\,,\!
  \label{eq:symmetryPropertyOfF}\\
  \rho_{ab}(x,y) &= \tau_{bc} \rho_{cd}(y,x) \tau_{da}
                  = (-1)^{i_a + \bar{i}_b}\!\rho_{\bar b\bar a}(y,x)\,,
\end{align}
\end{subequations}
where ${\bar a}=(\alpha,3-i_a)$ and $\bar{i}_a = 3-i_a$.
Note that in the definitions of $\rho$ and $F$ the operator product is not time ordered and cannot be directly evaluated in the path integral formalism.
However, expectation values of equal-time anticommutators of Gra\ss mann variables can be evaluated using the Bjorken-Johnson-Low theorem~\cite{Bjorken:1966jh,Johnson:1966se}.
For the spectral density function, one finds%
\footnote{The anticommutation relations are consistent with the anticommutation relations in the canonical quantization approach.}:
\begin{equation}
  \rho_{ab}(x,y)\bigr|_{x_{0}=y_{0}}
    = \myi\tau_{ab}\delta^{(d)}(\vector{x}-\vector{y})\,.
  \label{eq:equaltimeanticommutator}
\end{equation}
In the derivation of the equations of motion we made use of this anticommutation relation in the identity
\begin{equation}
\begin{split}
  \rho_{ab}(x,y)\del_{x_{0}}\sgnC{x_{0}-y_{0}}
  &= 2\rho_{ab}(x,y)\delta_{\mathcal{C}}(x_{0}-y_{0})\\
  &= 2\myi\tau_{ab}\delta_{\mathcal{C}}(x-y)\,.
\label{eq:dx0rho}
\end{split}
\end{equation}

\section{Conservation laws}
\label{app:ConsLaws}

For a theory of dynamics to be physically meaningful, it is crucial to respect the conservation laws prescribed by the symmetries present in nature.
For example, for a system that is closed with respect to the exchange of particles and energy with its surroundings, a theoretical description of its dynamics needs to reflect the conservation of particle number and energy irrespective of the approximation chosen.
Moreover, if the system relaxes to an equilibrated state in the long-time limit then this state is determined by the values of the conserved quantities in the initial state.
In the following, we show that the total particle number and energy are conserved at any order of truncation of the two-particle irrecducible (2PI) effective action.
To take into account other independent conserved quantities, a generalization of the 2PI effective action approach to $n$PI effective actions is required \cite{Berges:2004pu}.

\subsection{Particle number conservation}
\label{sec:NumberConservation}

According to Noether's theorem, the symmetry of the Lagrangian~\eq{lagrangiandesityforcomplexfields} with respect to global $U(1)$ transformations of the complex fields $\psi_{a}$ implies the conservation of the total particle number.
This can be shown as follows.

From the stationary condition~\eq{thestationarycondition}, one can construct the vanishing expression
\begin{equation}
  0 = -2\myi\tau_{ab}^{}
       \int_{y}
         \frac{\delta\varGamma[G]}{\delta G_{cb}(y,x)}
         G_{ca}(y,x) \,.
\end{equation}
Substituting the right-hand side of Eq.~\eqref{eq:varGamma} for $\varGamma$ into this expression, one finds
\begin{equation}
  \del_{x_{0}} n(x) + \vector{\nabla} \cdot \vector{j}(x)
  = -2\myi\tau_{ab}^{}
      \int_{y}
       \frac{\delta\varGamma_{\text{int}}[G]}{\delta G_{cb}(y,x)}
       G_{ca}(y,x)
   \label{eq:numberConservationGammaIntEquation}
\end{equation}
with
\begin{align}
  \varGamma_{\text{int}}[G]
  &= \varGamma[G]
     + \frac{\myi}{2}
       \Tr{G^{-1}_{0} G} \\
  n(x)
  &= \frac{\mathcal{N}}{2}
     \Bigl( 1 - G_{aa}(x,x) \Bigr) \\ 
\begin{split}
  \vector{j}(x)
  &= \frac{\myi\tau_{ab}}{2 m}
     \vector{\nabla}_{z}
     G_{ba}^{}(z,y) \biggl|^{}_{z=y=x}\,,
\end{split}
\end{align}
where sums over $a=(\alpha,i_a)$, $b=(\beta,i_b)$, and $c=(\gamma,i_c)$ are implied.
Like the underlying classical action $S$, \Eq{Sclass}, the effective action, and hence also $\varGamma_{\text{int}}$, is a singlet under $O(\mathcal{N})$ rotations and parameterized by the field $G$.
All functions that are singlet under $O(\mathcal{N})$ rotations can be built from the irreducible, i.\,e., in spin-index not factorizable, invariants $\text{tr}[G^n]$ with $n\leq\mathcal{N}$.
Here, $\text{tr}[\cdot]$ applies to the spin indices and does not include an integration over space-time coordinates, e.\,g., $\text{tr}[G^3] = G_{ab}(x,y) G_{bc}(y,z) G_{ca}(z,x)$.
Thus, the integrand of \Eq{numberConservationGammaIntEquation} is symmetric under an exchange of $\alpha$ and $\beta$.
$\tau_{ab}$ is antisymmetric under an exchange of $\alpha$ and $\beta$.
Hence, summing over $a$ and $b$ sets the right-hand side  of \Eq{numberConservationGammaIntEquation} to zero, and \Eq{numberConservationGammaIntEquation} becomes the continuity equation.
These symmetry considerations are independent of the specific approximation to $\varGamma_{\text{int}}$.
Thus, for a closed system, the total particle number is conserved at any order of truncation of the 2PI effective action.

\subsection{Energy conservation}
\label{sec:EnergyConservation}

For time independent interactions and external potential, the Lagrangian is time translation invariant.
This implies energy conservation.
Here, we consider the invariance under general translations in continuous space and time that vanish at the boundary, $x^\mu\to x^\mu+\epsilon^\mu(x)$, where $\epsilon^\mu(x)$ is a time- and space-dependent infinitesimal $(d+1)$-vector.
To leading order in $\epsilon$, the Green function transforms under these translations as
   $G_{ab}(x,y) \to G_{ab}(x,y)
                    + \epsilon^\nu(x)\partial^x_\nu G_{ab}(x,y)
                    + \epsilon^\nu(y)\partial^y_\nu G_{ab}(x,y)$,
where $\partial^x_\nu=\partial/\partial x^\nu$.
One can show that under these transformations the variation of the 2PI effective action $\varGamma$ can be written as
  $\varGamma[G] \to \varGamma[G] + \delta\varGamma[G]$,
with
\begin{eqnarray}
  \delta\varGamma[G] = \int_x T^{\mu\nu}(x)\,\partial^x_\mu \epsilon_\nu(x)\,.
\label{eq:deltaGamma}
\end{eqnarray}
Since, by virtue of the stationarity condition (\ref{eq:thestationarycondition}), the variation $\delta\varGamma$ vanishes for all solutions of the equation of motion for $G$, an integration by parts shows that $T^{\mu\nu}$ is the conserved Noether current for the time-space-translations:
\begin{eqnarray}
  \delta\varGamma[G] = -\int_x \epsilon_\nu(x)\,\partial^x_\mu T^{\mu\nu}(x)
                     = 0\,.
\label{eq:EMTconservation}
\end{eqnarray}
$T^{\mu\nu}(x)$ is identified as the energy-momentum tensor, and the conservation law for total energy is expressed as $\partial^x_\mu T^{\mu0}(x)=0$ or $\partial_t\int \mathrm{d}^{(d)}x\,T^{00}(t,\vec{x})=0$.

As sketched in the following, the energy momentum tensor can be calculated by standard techniques for any truncation of the 2PI effective action.
The energy density is found to be
\begin{equation}
  T_{00}(x)
  =\varepsilon_{\text{kin}}(x) + \varepsilon_{\text{pot}}(x)
\label{eq:EMT2PI1N_00}
\end{equation}
with
\begin{align}
  \varepsilon_{\text{kin}}(x)
  &= \int_{y}
        \delta(x-y)
        H^{\text{1B}}_{\alpha\alpha}(x)
        n^{}_{\alpha\alpha}(x,y)\,,
\label{eq:totalKineticEnergyDensity}
\\
\begin{split} \label{eq:totalPotentialEnergyDensity}
  \varepsilon_{\text{pot}}(x)
  &= \frac{\lambda^{}_{\alpha\beta}}{2\mathcal{N}}
         n^{}_{\alpha}(x)
         \bigl( 1 - n^{}_{\beta}(x) \bigr)
    -\frac{1}{2}
     I^{}_{\alpha\alpha}(x,x) \,,
\end{split}
\end{align}
where sums over $\alpha$ and $\beta$ are implied.
The conserved total energy is then given as the spatial integral over the above density.

We close this appendix with a sketch of the derivation of the energy momentum tensor for an ultracold Fermi gas in $d$ spatial dimensions, described by the 2PI effective action in next-to-leading order (NLO) $1/\mathcal{N}$ approximation.
To this end, the variation of the effective action $\varGamma$ under space-time translations is split into one-loop and higher order terms,
\begin{equation}
  \delta\varGamma[G]
    = -\frac{\myi}{2} \delta\left(
      \Tr{\ln{G^{-1}_{}} + G^{-1}_{0} G} \right)
      +\delta\varGamma^{}_{2}[G] \,.
\label{eq:variationOfVargamma}
\end{equation}
To obtain the contribution to $T^{\mu\nu}$ arising from the one-loop term, we use \Eq{deltaGamma} and observe that
the term $\delta\Tr{\ln{G^{-1}_{}}}$ in Eq.~\eqref{eq:variationOfVargamma} does not contribute to $T^{\mu\nu}$.
The term $\Tr{G^{-1}_{0} G}$ can be written as
\begin{equation}
\begin{split} \label{eq:termG^-1_0G}
 &\delta
  \left(
    -\frac{\myi}{2}
     \Tr{G^{-1}_{0}(x,y)G(y,x)}
  \right) \\
 &= -\frac{\myi}{2}
    \int_{xy}
    G^{-1}_{0,ab}(x,y)
    \delta G^{}_{ba}(y,x)
\end{split}
\end{equation}
with 
(the potential is assumed to vanish, $V_{\text{ext}, \alpha\beta}=0$)
\begin{align}
  \myi G^{-1}_{0,ab}(x,y)
  &= \delta(x-y)
     \Bigl(
       \myi\tau_{ab}\del^{y}_{0}
       + \delta_{ab}
         \frac{1}{2m}\del^{y}_{k}\del^{y}_{k}
     \Bigr)\,,
\\
\begin{split}
  \delta G^{}_{ab}(x,y)
  &= \Bigl(
       \varepsilon^{\lambda}(x) \del^{x}_{\lambda} 
       + \varepsilon^{\lambda}(y) \del^{y}_{\lambda}
     \Bigr)
     G^{}_{ab}(x,y)\,.
\end{split}
\end{align}
Integration by parts and the identity
  $\int_x \del^{y}_{k}[\delta(x-y)G(y,x)]
   = \int_x \delta(x-y) (\del^{x}_{k} + \del^{y}_{k}) G(y,x)$
allows to rewrite the r.\,h.\,s.\ of Eq.~\eqref{eq:termG^-1_0G} in the form of \Eq{deltaGamma}.
The resulting contribution to the energy density $T_{00}(x)$ is
\begin{equation}
  \varepsilon_{\text{kin}}(x)
   = -\frac{1}{\n}
      \int_{y}
        \delta(x-y)
        \delta_{ab}
        H^{\text{1B}}_{\alpha\beta}(x)
        G^{}_{ba}(x,y)\,.
\label{eq:kineticEnergyDensity}
\end{equation}

To obtain the contribution from the term $\varGamma^{}_{2}$ to the energy-momentum tensor, it is more convenient to use a method known from field theory on curved space-time, where, for a space-time-dependent metric $g^{\mu\nu}_{}$, the energy-momentum tensor is given by~\cite{Misner1973a}
\begin{equation}
  T_{\mu\nu}(x)
  = \frac{2}{\sqrt{-g(x)}}
    \frac{\delta\varGamma[G,g^{\mu\nu}_{}]}{\delta g^{\mu\nu}_{}(x)}
\end{equation}
with $\sqrt{-g(x)}$ denoting the square root of minus the determinant of $g^{}_{\mu\nu}(x)$.
We apply this to  $\varGamma^{}_{2}$ in NLO of the $1/\mathcal{N}$ expansion, cf.\ Eqs.~(\ref{eq:varGamma21rNsummation}--\ref{eq:definitionOfGamma2NLO}), with $\int_x = \int\text{d}x\sqrt{-g(x)}$.
Making use of the identity
  $\delta\sqrt{-g(x)}/\delta g^{\mu\nu}_{}(y)
   = -\sqrt{-g(x)} g^{}_{\mu\nu}(x) \delta(x-y)/2$
in flat space with $g_{\mu\nu}(x)=\text{diag}\{1,-1,-1,-1\}$,
one finds
\begin{align}
  \frac{2}{\sqrt{-g(x)}}
     \frac{\delta\varGamma_{2}^{\text{LO}}}
          {\delta g^{\mu\nu}_{}(x)}
  &= g^{}_{\mu\nu}(x)
     \frac{\lambda_{\alpha\beta}}{8\mathcal{N}}
     G^{}_{aa}(x,x) G^{}_{bb}(x,x)\,,
  \label{eq:LO_energyMomentumTensorContribution}
\\
  \frac{2}{\sqrt{-g(x)}}
     \frac{\delta\varGamma_{2}^{\text{NLO}}}
          {\delta g^{\mu\nu}_{}(x)}
  &= -\frac{1}{2}
     g^{}_{\mu\nu}(x)
     I_{\alpha\alpha}(x,x) \,,
  \label{eq:NLO_energyMomentumTensorContribution}
\end{align}
where the function $I^{}_{\alpha\beta}(x,y)$ is given by Eq.~\eqref{eq:DefitionResummedBubbleChainI}.
Finally, using
  $G^{}_{aa}(x,x)
   = F^{}_{aa}(x,x)
   =  1 - 2n_{\alpha}(x) $
allows to rewrite Eqs.~\eqref{eq:kineticEnergyDensity}, \eqref{eq:LO_energyMomentumTensorContribution} and \eqref{eq:NLO_energyMomentumTensorContribution} to obtain the results \eqref{eq:totalKineticEnergyDensity} and \eqref{eq:totalPotentialEnergyDensity}, where also the vacuum energies are removed.

\end{appendix}

} 

\end{document}